\documentclass[epj]{svjour}
\usepackage{amsmath,amssymb,graphicx}
\usepackage[stable]{footmisc}
\usepackage{color}
\usepackage{subfig}
\usepackage{wrapfig,lipsum}
\usepackage{tabularx}
\usepackage{algorithm2e}

\pdfoutput=1

\newcommand{\nc}{\newcommand}
\nc{\postscript}[2]
{\setlength{\epsfxsize}{#2\hsize}\centerline{\epsfbox{#1}}}
\nc{\non}{\nonumber}
\nc{\hc}{\hbox {h.c.}} \nc{\re}{\hbox {Re}} 
\nc{\mev}{\hbox {MeV}} \nc{\gev}{\;\hbox {GeV}} \nc{\tev}{\;\hbox {TeV}}
\def\lsim{\mathrel{\raise.3ex\hbox{$<$\kern-.75em\lower1ex\hbox{$\sim$}}}}
\def\gsim{\mathrel{\raise.3ex\hbox{$>$\kern-.75em\lower1ex\hbox{$\sim$}}}}

\nc{\etal}{{\it et al.}}
\nc{\Lsp}{\;\;\;\;\;\;\;\;\;\;}  \nc{\LLLsp}{\lspace \lspace}
\nc{\lsp}{\;\;\;\;\;\;}
\nc{\spac}{\;\;\;}
\nc{\noi}{\noindent}
\nc{\beq}{\begin{equation}}   \nc{\eeq}{\end{equation}}
\nc{\bea}{\begin{eqnarray}}   \nc{\eea}{\end{eqnarray}}
\nc{\baa}{\begin{array}}      \nc{\eaa}{\end{array}}
\nc{\bit}{\begin{itemize}}    \nc{\eit}{\end{itemize}}
\nc{\ben}{\begin{enumerate}}  \nc{\een}{\end{enumerate}}
\nc{\bce}{\begin{center}}     \nc{\ece}{\end{center}}

\def\sq2{\sqrt{2}}

\def\ph{\varphi}

\def\m4{m^4(\ph)}
\def\mn2{m_n^2}

\def\v5{V^{(5)}}

\def\baa{\begin{array}}
\def\eaa{\end{array}}

\begin{document}
\title{Electroweak Stability and Discovery Luminosities for New Physics}
\author{Kerem Canko{\c c}ak\inst{1} \and Durmu{\c s}  Demir\inst{2}\and Canan Karahan\inst{1}  \and Sercan {\c S}en\inst{3} 
}                     
%
%
\institute{Physics Engineering Department, {\.I}stanbul Technical University, 34469 Maslak, {\.I}stanbul, Turkey \and Faculty of Engineering and Natural Sciences, Sabanc{\i} University, 34956 Tuzla, {\.I}stanbul, Turkey \and Physics Engineering Department, Hacettepe University, 06800 Beytepe, Ankara, Turkey}
\date{Received: date / Revised version: date}
%
\abstract{
What is the luminosity needed for discovering new physics if the electroweak scale is to remain stable? In this work we study this question, with the pertinent example of a real singlet scalar which couples to the Higgs field at the renormalizable level. Observing that the electroweak scale remains stable if the two scalars couple in a see-sawic fashion through a mass-degeneracy- driven unification linkup among quartic couplings at a given scale, we show by detailed simulation studies of the $pp\to ({\rm singlet\ scalar}) \to Z Z \to 4\ell$ channel that the HL-LHC, which is expected to deliver an integrated luminosity of $3\ {\rm ab^{-1}}$, has no significant excess of signal over the background in the 800 -- 2000 GeV mass range. The FCC-hh, on the other hand, can discover scalars up to a mass of 870 GeV with an integrated luminosity $20\ {\rm ab^{-1}}$. Observation at $3\sigma$ (discovery at $5\sigma$) of a new scalar with a minimum mass 800 GeV requires at least $2\ {\rm ab^{-1}}$ ($5.2\ {\rm ab^{-1}}$) integrated luminosity, showing that the new physics that does not destabilize the electroweak scale is accessible only at very high luminosities, and can be tested already in the early stages of the FCC-hh operation period.}

\PACS{
      {12.60-i}{Models beyond the standard model}   \and
      {12.15.Lk}{Electroweak radiative corrections}
     } 

\maketitle
\section{Introduction}
The standard model of elementary particles (SM), experimentally completed by the discovery of the Higgs boson at the ATLAS and CMS  \cite{higgs}, has shown excellent agreement with all  the available data so far \cite{sm-sofar}. The TeV domain seems to be devoid of any new particles beyond the SM spectrum \cite{exotica}.  The SM seems thus to continue to hold good up to energies well above a TeV. 

This experimental fact poses a big challenge. The reason is that there are all sort of astrophysical (dark matter, dark photon), cosmological (dark energy, inflation) and other (neutrino masses, flavor, unification, $\cdots$) indications that the SM has to be extended by new fields  beyond the TeV \cite{csaki}. These fields set the so-called physics-beyond-the-SM (BSM) sector.

The problem with the BSM phenomena is that they generally drag the SM towards their scales. Indeed, if the SM+BSM is valid up to a scale $\Lambda$ and if the Higgs boson couples with strength $\lambda_{h-BSM}$ to BSM fields of masses $M_{BSM}$ then its mass $m_h$ receives the correction
\begin{eqnarray}
\label{higgs0}
\delta m_h^2 = c_h \Lambda^2 +  {\tilde{c}}_{h} \lambda_{h-BSM}  M^2_{BSM} \log \frac{M^2_{BSM}}{\Lambda^2}
\end{eqnarray}
where $c_h$ and ${\tilde{c}}_{h}$ are ${\mathcal{O}}(10^{-2})$ loop factors, and the Higgs-BSM coupling $\lambda_{h-BSM}$ changes from portal to portal in view of the SM-BSM coupling channel:
\begin{equation}
\lambda_{h-BSM} \propto \left\{\begin{array}{ll}
\lambda_{HS} & {\rm \;\;\;  \!\!\! for \ Higgs \ portal}: \lambda_{HS} H^\dagger H S^\dagger S\\ \lambda^2_{Z^\prime B}&{\rm \;\;\; \!\!\! for \ hypercharge \ portal}: \lambda^2_{Z^\prime B} Z^\prime_{\mu\nu} B^{\mu\nu}\\ \lambda^2_{LN}&{\rm \;\;\;  \!\!\! for \ neutrino \ portal}: \lambda_{LN} \overline{L}H N + {\rm h.c.}
\end{array} \right.
\label{portals}
\end{equation}
The lesson from the LHC experiments is that the UV completion that eliminates the $\Lambda^2$ term in (\ref{higgs0}) must do it with a sufficiently {\it small} $\lambda_{h-BSM}$ in order to keep $\delta m_h^2/m_h^2$ sufficiently small. It is in this sense that the known completions of the SM (supersymmetry, extra dimensions, compositeness, and their hybrids) are sidelined because their BSM sectors (superpartners in supersymmetry, Kaluza-Klein levels in extra dimensions, and technifermions in compositeness)  necessitate $\lambda_{h-BSM}\simeq \lambda_{SM}$, where $\lambda_{SM}$ is a typical SM coupling. It turns out that heavier the BSM larger the shift in the Higgs boson mass, and stronger the exclusion limits on the known SM completions. In view of this LHC lesson, we do in the present work three things:
\begin{enumerate} 
\item In Sec. II below we discuss in detail how an 
{\it LHC-favored SM completion} must be, and give symmergence \cite{demir2017,demir2019} as a likely realization in which $c_h \Lambda^2$ part is eliminated (actually transmuted to curvature) by keeping  ${\tilde{c}}_h$ sufficiently {\it small}. Therein we analyze how ${\tilde{c}}_h$ can be structured, and construct a linkup scheme in which the seesawic coupling in \cite{demir2019} is reproduced.

\item In Sec. III and Sec. IV, without loss of generality, focus and goal, we specialize to a BSM sector in which only  a real singlet scalar interacts with the SM Higgs boson, and reveal there the implications of the see-sawic couplings of Sec. II in view of the electroweak stability. (In general, one should analyze all three portals in (\ref{portals}) but for showing the stability of the electroweak scale against the logarithmic corrections in (\ref{higgs0}) it proves beneficial to focus on a singlet scalar $S$ as the worst case because $S$ itself is a power-law UV sensitive field.)  

\item Sec. V we study the collider phenomenology of the real SM-singlet scalar in Sec. III, with emphasis on electroweak stability. There we attempt to answer this key question:
\begin{eqnarray}
&&{\rm What\ energy\ and\ luminosity\ does\ it\ take\ to\ discover}\nonumber\\ &&{\rm a\ singlet \ scalar}\ {\rm if\ the\ electroweak\ scale\ is\ to\ remain}\nonumber\\ &&{\rm stable?}
\label{question}
\end{eqnarray}
To answer this question we perform a detailed analysis of the decay and production channels associated with the singlet scalar, and determine explicitly discovery luminosities at the HL-LHC and FCC-hh colliders via the low-background 4-lepton signal. Our analysis in Sec. V shows how LHC-favored SM completions like symmergence can reveal themselves at high luminosities. 
\end{enumerate}
In Sec. VI we conclude.

\section{LHC-Favored SM Completion}
\label{sect:favored}
If the LHC results \cite{higgs,exotica} have taught us anything it is that the SM can be extended with a BSM sector (facilitating neutrino masses, baryogenesis, dark matter, inflation, $\dots$) naturally if SM-BSM couplings (like $\lambda_{h-BSM}$ in $\delta m_h^2$) are allowed to be {\it small} ($\lambda_{h-BSM}\ll \lambda_{SM}$) and if the power-law UV sensitivities (like $c_h \Lambda^2$ in $\delta m_h^2$) are properly {\it nullified}. There are thus two sides of the problem, and both sides need to be properly addressed. 

The nullification of the power-law UV sensitivities is achieved ordinarily in all the known UV completions. They are eradicated in supersymmetry, downsized to TeV in extra dimensions, and turned into curvature in symmergence (emergence of gravity in a way restoring gauge symmetries broken explicitly by the UV cutoff $\Lambda$). These UV sensitivities (like $c_h \Lambda^2$ in $\delta m_h^2$ in equation (\ref{higgs0}) above) are thus far from destabilizing the electroweak scale. 
  
The logarithmic UV sensitivities (like the $\lambda_{h-BSM}$ term in $\delta m_h^2$ in equation (\ref{higgs0}) above) are highly nontrivial in that the known SM completions have $\lambda_{SM-BSM}\simeq \lambda_{SM}$, and never get into the hierarchic regime $\lambda_{SM-BSM}\ll \lambda_{SM}$. This is a serious problem because  heavier the BSM larger the logarithmic part of $\delta m_h^2$ since it is proportional to $\lambda_{h-BSM} M_{BSM}^2$ and since all known UV completions necessitate $\lambda_{SM-BSM}\simeq \lambda_{SM}$. The logarithmic part poses  thus as a serious hierarchy problem (causing the so-called little hierarchy problem \cite{little-hierarchy}). It is in this sense that the LHC results started sidelining known UV completions like supersymmetry, extra dimensions and compositeness. They point to new SM completions which can work with $\lambda_{SM\!-\!BSM}\\
\!\!\ll\!\! \lambda_{SM}$ instead of $\lambda_{SM-BSM}\simeq \lambda_{SM}$ (characteristic to  supersymmetry, extra dimensions and compositeness). 
 
\subsection{Symmergence as the LHC-Favored UV Completion}
The focus in the present work is on the logarithmic quantum corrections, that is, on the little hierarchy problem posed, for instance, by the logarithmic part of $\delta m_h^2$ in (\ref{higgs0}). In other words, one gives weight to logarithmic corrections with the assumption that there exists some UV completion that does away with the power-law UV sensitivities \cite{veltman}. It is, nevertheless, necessary to have a concrete UV completion in mind even if it may not phenomenologically be directly relevant for the analysis of the effects of the logarithmic corrections. There is currently one such UV completion: 
Symmergent gravity \cite{demir2017,demir2019,demir2016}. To have an idea what symmergence is and is not one notes first that, in the presence of the UV momentum cutoff $\Lambda$, the photon and the gluon acquire masses (with zero logarithmic parts)
\begin{eqnarray}
\label{gam-g}
\delta m_{\gamma,g}^2 = c_{\gamma,g} \Lambda^2 
\end{eqnarray}
which completely  destruct the SM by breaking the color and the electric charge (CCB) \cite{demir2016,ccb}. (In a broader sense, these are gauge 
anomalies.) This quantum gauge invariance breaking can be prevented if the quadratic corrections $c_g \Lambda^2$ and $c_\gamma \Lambda^2$ are somehow neutralized. 

The gauge boson mass corrections (\ref{gam-g}) are physical. They cannot be altered by any means simply because the cutoff $\Lambda$ is physical. It is physical because it refers to physical phenomena  beyond the SM (like gravity and possible BSM physics). If gravity were absent, if there were no BSM physics, if it were just the SM alone then all loop integrals would turn into cutoff regularization integrals \cite{cutoff} with arbitrary cutoff $\Lambda$, and one would then be able to eradicate all power-law corrections (like (\ref{gam-g}) and (\ref{higgs0})) simply by switching to dimensional regularization \cite{dim-reg}. The essence of the problem is that gravity and BSM physics (needed for dark matter, strong CP, baryogenesis, and various other phenomena) introduce new physical scales like $\Lambda$, and neutralization of the gauge symmetry-breaking corrections like (\ref{gam-g}) necessitates a new mechanism that complies with the SM as well as the BSM and gravity. 

In regard to neutralization of the loop-induced gauge boson masses in (\ref{gam-g}), symmergence is a mechanism resulting from the observation that it should be possible to set up a covariance relation (just like the usual general covariance between the flat metric and the curved metric) between $\Lambda^2$ (which explicitly breaks Poincare invariance of the flat spacetime on which QFTs like the SM are based) and spacetime curvature (which explicitly breaks Poincare invariance as a built-in feature of the curved spacetime). This extended covariance, which rests on the Poincare affinity between $\Lambda^2$ and curvature, is guided by gauge invariance (as discussed in detail in \cite{demir2019}).  In other words, the proposed covariance between $\Lambda^2$ and curvature must be able to restore gauge invariance by  transmuting  the loop-induced gauge boson masses in (\ref{gam-g}) appropriately. This transmutation mechanism  leads to symmergence \cite{demir2019,demir2016}, with the following salient features:
\begin{enumerate}
    \item It predicts existence of a BSM sector (of neutrinos, dark matter, $\dots$, and maybe more),
    
    \item It kills the gauge boson mass term (\ref{gam-g}) (restores gauge invariance) by transmuting $\Lambda^2$ into affine curvature,
    
    \item It converts $c_h \Lambda^2$ part of (\ref{higgs0}) into Higgs-curvature coupling,
    
    \item It converts $c_S \Lambda^2$ part of the corrections $\delta m_S^2$ to the BSM scalar masses into $S$-curvature couplings,
    
    \item It leads to Einstein gravity, and finally,
    
    \item It results in dimensionally-regularized SM+BSM in the curved spacetime.
\end{enumerate}
What is left untouched by symmergence is the logarithmic part of (\ref{higgs0}). It is left untouched because $\log \Lambda^2$ does not break Poincare invariance (it always leads to multiplicative corrections). It remains as a physical contribution to the Higgs boson mass. This remnant contribution might give the impression that symmergence makes no real progress concerning the electroweak stability. No! Actually it makes a pivotal progress. It does because it leaves couplings between the SM and the BSM unconstrained, that is,  free to take any perturbative value, even the zero! (This feature rests on the fact that the gravitational scale is set by the supertrace of the SM+BSM mass-squareds, with no necessity of any couplings between the SM and the BSM. In other words, the SM and BSM do not have to interact.) This means that the SM-BSM couplings can be small enough (compared to the SM couplings) to suppress logarithmic corrections. This SM-BSM coupling scheme is something specific to symmergence. Indeed, contrastively, 
\begin{eqnarray}
\label{normal-coupl}
\lambda_{S\!M\!\!-\!\!B\!S\!M} \! \simeq \! \lambda_{S\!M}\! \Longrightarrow \! 
{\rm only\ light\ BSMs\ with}\
m_{B\!S\!M}\!\gtrsim \! m_h
\end{eqnarray}
are allowed in supersymmetry, extra dimensions and compositeness \cite{csaki} whereas 
\begin{eqnarray}
\label{symm-coupl}
\lambda_{S\!M\!\!-\!\!B\!S\!M}\! \ll \!\lambda_{SM}\! \Longrightarrow \!{\rm heavy\ BSMs\ with}\
m_{B\!S\!M} \gg m_h\ 
\end{eqnarray}
are allowed in symmergence \cite{demir2016,demir2019}. It thus turns out that the LHC can exclude sparticles (the BSM of supersymmetry), Kaluza-Klein levels (the BSM of extra dimensions) and technifermions (the BSM of compositeness) but not the BSM of the symmergence! In fact, the Higgs mass correction in (\ref{higgs0}) remains within the LHC bounds if the SM-BSM couplings obey the bound
\begin{eqnarray}
\label{seesawic0}
\lambda_{h-BSM} \lesssim \frac{m_H^2}{m^2_{BSM}}
\end{eqnarray}
which is a seesawic relation between the mass parameters of the Higgs field $H$ ($m_h^2= - 2 m_H^2$ in the SM) and the BSM fields. This seesawic coupling scheme, ensuring that heavier the BSM smaller its couplings to the SM, gives way to a novel approach to collider and other searches for the BSM physics. 

It should be noted that stabilization of the SM against $\Lambda^2$ corrections has been approached via various mechanisms. They include classical conformal invariance \cite{scale},  twin Higgs  \cite{twin-higgs}, cosmological relaxation \cite{relax-Higgs}, Higgs frame \cite{conformal-frame}, gravitational relaxation \cite{relax-grav}, large copies of the SM spectrum \cite{large-copies-SM}, and subtraction of $\Lambda^2$ terms  \cite{quad}. The subtraction method \cite{quad} is particularly relevant in that it subtracts (absorbs into critical surface) $\Lambda^2$ terms, retains only $\log\Lambda$ terms (dimensional regularization), predicts no BSM sector, and leaves out gravity entirely. Symmergence \cite{demir2019}, on the other hand, identifies $\Lambda^2$ terms with curvature by their Poincare affinity, retains only $\log\Lambda$ terms (dimensional regularization), predicts existence of a BSM sector, and makes gravity emerge upon the SM+BSM. This clear distinction between the two approaches shows that symmergence could indeed be a factual UV completion of the SM.

\subsection{Mass-Degeneracy-Driven Unification Linkup}
The seesawic scheme in (\ref{seesawic0}), an empirical relation based on the freedom in (\ref{symm-coupl}) enabled by symmergence, is a just-so relation. The thing is that it works. The problem is that  there is no obvious symmetry principle that supports it. (This can be seen from the symmetry structures of the two-Higgs doublet models  \cite{haber,haber2}.)  To this end, it proves convenient to specialize to a real BSM scalar $S$  of mass $m_S$. (The remaining BSM fields do not have to couple to the SM; they can stay with zero couplings to the SM fields. But if any of them couples to the SM Higgs its effects must be taken into account in view of the portals in (\ref{portals}).)  The scalar $S$ couples to the SM Higgs field via the potential 
\bea\label{eq:pot}
V_{H\!S\!}\!=\!m_{\!H\!}^2H^{\dagger}\!H\!\!+\!\!\lambda_H\!(H^{\dagger}\!H)^2\!\!+\!\!\frac{m_S^2}{2}S^2\!+\!\!\frac{\lambda_S}{4}S^4\!\!+\!\!\frac{\lambda_{H\!S\!}}{2}H^{\dagger}\!H\!S^2
\eea
whose boundedness from below necessitates
\begin{eqnarray}
\label{conds}
\lambda_H > 0\;,\;\; \lambda_S > 0\;,\;\;  16 \lambda_H \lambda_S - \lambda_{HS}^2 > 0
\end{eqnarray}
as primary constraints on dimensionless couplings. 

It follows from the potential (\ref{eq:pot}) that the $\log \Lambda$ sensitivities (translated into dimensional regularization via the formal equivalence $\log \Lambda^2=1/\epsilon -\gamma_E+1+\log 4\pi Q^2$)  lead to non-trivial $\overline{MS}$ corrections to the  Higgs condensation parameter $m_H^2$ ($m_h^2=-2 m_H^2$ in the SM)\begin{eqnarray}
\label{corr-1}
\delta m_H^2 = c_H \lambda_{HS} m_S^2 \log \frac{m_S^2}{Q^2}
\end{eqnarray}
as follows from (\ref{higgs0}) (whose the quadratic part goes into curvature terms via symmergence \cite{demir2019}). It is clear that larger the $m_S$ larger the $\delta m_H^2$ and stronger the destabilization of the electroweak scale. The question is clear: How to prevent this destabilization of the electroweak scale? This is a profound question. And its answer is both obvious and obscure. It is obvious in that $|\lambda_{HS}|$ must be just {\it small} (as in (\ref{seesawic0}) above) to start with since loop corrections to $\lambda_{HS}$ are proportional to $\lambda_{HS}$ itself (it remains small if it is small). It is, on the other hand, {\it obscure} in that there is no obvious selection rule or symmetry that can ensure the requisite smallness \cite{haber,haber2}. Its dimensionless nature disfavors also dynamical mechanisms like Giudice-Masiero mechanism \cite{g-m} because a change like $\lambda_{HS} \rightarrow H^\dagger H/S^2$ would simply mean  killing the coupling between $H$ and $S$. In the face of this impasse, a reasonable way to follow would be imposition of a judicious relationship among model parameters, with stabilization under the renormalization group flow.  Indeed, symmergence, which sets $\lambda_{HS}$ free (as opposed to the known completions which require $\lambda_{HS}$ to remain close to the Higgs quartic coupling $\lambda_H$ as in (\ref{normal-coupl}) above), opens room for a mechanism in which  $\lambda_{HS}$ can be linked to other model parameters in a way that keeps $\delta m_H^2$ under control. In this respect, the seesawic coupling in (\ref{seesawic0}), which takes the particular form 
\begin{eqnarray}
\label{seesawic2}
\lambda_{HS} \propto \lambda_H \frac{m_H^2}{m_S^2} 
\end{eqnarray}
for a BSM scalar $S$, possesses the right structure to keep $\delta m_H^2$ in (\ref{corr-1}) below $m_H^2$. It is clear that, in view of the bound (\ref{seesawic0}), this coupling corresponds to the largest allowed values in that perfectly allowed smaller values result in feebler signals which are hard to detect.

The seesawic structure in (\ref{seesawic2}) relates $\lambda_{HS}$ to the field masses. This means that 
for stabilizing the Higgs mass (suppressing $\delta m_H^2$ in (\ref{corr-1})) the parameters in the potential must somehow enjoy a mass-dependent relationship beyond the energy conditions in (\ref{conds}). To this end, taking into account the perturbativity, we introduce a Mass-Degeneracy-Driven Unification (MDDU) linkup of the form (at a given scale $Q=Q_0$)
\begin{eqnarray}
\label{link}
\lim_{m_H(Q_0)\rightarrow m_S(Q_0)} \lambda_H(Q_0) \!=\! \lambda_S(Q_0) =|\lambda_{HS}(Q_0)|
\end{eqnarray}
which proves useful as it possesses the particular solution
 \begin{eqnarray}
\lambda_S(Q_0)\! =\! \lambda_H(Q_0), \hspace{0.2cm}    \lambda_{HS}(Q_0) = \frac{2 \lambda_{H}(Q_0)}{\frac{m_H^2(Q_0)}{m_S^2(Q_0)} + \frac{m_S^2(Q_0)}{m_H^2(Q_0)}}
    \label{linkup}
    \end{eqnarray}
according to which $\lambda_{HS}(Q_0)$ reduces to the seesawic structure in (\ref{seesawic2}) for  $m_S(Q_0)\gg m_H(Q_0)$, and smoothly covers the opposite limit of $m_S(Q_0)\ll m_H(Q_0)$. It is clear the MDDU linkup (\ref{link}) introduces a mass-dependent correlation among the quartic couplings in (\ref{eq:pot}). It is special point in the parameter space (rather than a symmetry principle \cite{haber,haber2}).  
The particular MDDU scheme in (\ref{linkup}) can be generalized to other portals in (\ref{portals}) by simply replacing $m_S^2$ with $M_{BSM}^2$, where $M_{BSM}=M_{Z^\prime}$ or $M_{BSM}=M_{N}$. This replacement rule can be useful for analyzing more general SM-BSM interactions.  
 
It is only with UV completions like symmergence that it $\lambda_{h-BSM}$ gets loosened from the SM couplings (in view of the fixture in (\ref{normal-coupl})), and it is with the MDDU in (\ref{link}) that seesawic couplings like (\ref{seesawic0}) or (\ref{seesawic2}) become possible. 

The changes in the MDDU linkup in (\ref{link}) and (\ref{linkup}) due to RGE flow of the parameters are analyzed in detail in the recent paper \cite{cemle}. It is shown there that within the perturbative regime the MDDU relation is quite robust. Indeed, $\lambda_{HS}(Q)$ is small, it remains small as ensured by its RGE, and the $\lambda_S(Q)$ and $\lambda_H(Q)$ in turn remain essentially unchanged.  Furthermore,  mass of Higgs  boson remains unaffected by the heavy scalar $S$ not at a specific scale $Q_0$ but at all scales from electroweak one to Planck one thanks to the smallness of $\lambda_{HS}(Q)$, as ensured the MDDU linkup. This shows that the MDDU works to ensure stability of the electroweak scale.

\section{The Model}
In this section we analyze effects of an SM-singlet real scalar $S$ on the electroweak stability. It certainly is possible to consider a wider BSM sector and include all three types of the portals in (\ref{portals}). Nevertheless, the Higgs portal suffices for demonstrating the stability of the  electroweak scale under the seesawic coupling in (\ref{linkup}). 

In view of the question (\ref{question}), the most general, renormalizable, symmetric Lagrangian density 
extending the SM with a real singlet scalar field $S$ is given by \cite{singlets}
\bea
\mathcal{L}_{HS}&=&(D_{\mu}H)^{\dagger}D^{\mu}H+\frac{1}{2}\partial_{\mu}S\partial^{\mu}S-V_{H\!S},
\eea
where $V(H,S)$ is the potential energy density in (\ref{eq:pot}), and $H$ is the usual SM Higgs doublet 
\bea
\label{Higgs-doublet}
H=\frac{1}{\sqrt{2}}\left(\baa{c}
\phi_1+i\phi_2 \\
\upsilon_H+h+i\phi_0
\eaa
\right).
\eea
with the Higgs boson $h$ remaining as a CP-even scalar after the Goldstone bosons $\phi_i$ are swallowed as longitudinal components of the $W$ and $Z$ bosons. Indeed,  for $\lambda_H>0$ and $\lambda_S>0$, the potential gets bounded from below  and the minimum of the potential breaks the electroweak symmetry spontaneously via the Higgs vacuum expectation value (VEV) $\upsilon_H\neq0$. If the scalar $S$ is not inert (see for instance \cite{non-inert,non-inert2}), that is, if it gets a VEV $\upsilon_S\neq 0$ then the minimum of the potential (\ref{eq:pot}) occurs at 
\bea
\upsilon_H^2\!=\! \frac{4\lambda_S m_H^2\!-\!2\lambda_{HS} m_S^2}{\lambda_{HS}^2-4\lambda_{H}\lambda_S},\!\!\!\!\!\qquad\!\!\!\!\! \upsilon_S^2\!=\!\frac{4\lambda_H m_S^2\!-\!2\lambda_{HS} m_H^2}{\lambda_{HS}^2-4\lambda_{H}\lambda_S}
\label{minimum}
\eea
with the singlet boson $s$ defined as $S=\upsilon_S + s$ in parallel to (\ref{Higgs-doublet}). 

In the vicinity of the vacuum (\ref{minimum}), the mass-squared matrix of the $h$ and $s$ bosons 
\bea
M^2=\left(\baa{cc}
2\lambda_H\upsilon_H^2 & \frac{1}{2} \lambda_{HS}\upsilon_H\upsilon_S \\
\frac{1}{2} \lambda_{HS} \upsilon_H\upsilon_S & 2\lambda_S\upsilon_S^2
\eaa
\right)
\eea
assume two eigenvalues 
\bea\label{eq:mh2msig2}
m_{h_1}^2\!\!&=&\!\!\lambda_H\upsilon_H^2\!\!+\!
\lambda_S\upsilon_{S}^2
\!-\!\sqrt{\left(\lambda_S\upsilon_{S}^2\!-\!\lambda_H\upsilon_H^2\right)^2
\!+\! \frac{1}{4}\lambda_{HS}^2 \upsilon_{S}^2\upsilon_{H}^2}, \non\\
m_{h_2}^2\!\!&=&\!\!\lambda_H\upsilon_H^2\!\!+\!
\lambda_S\upsilon_{S}^2
\!+\!\sqrt{\left(\lambda_S\upsilon_{S}^2\!-\! \lambda_H\upsilon_H^2\right)^2
\!+\!\frac{1}{4} \lambda_{HS}^2 \upsilon_{S}^2\upsilon_{H}^2}.\non\\
\eea
corresponding to the two physical eigenstates $h_1$ (which should be identified with the scalar boson observed at the LHC \cite{higgs}) and $h_2$ (the extra scalar boson under search at the LHC and to be searched for at future colliders like the FCC). The key parameter is their mixing angle 
\bea\label{eq:tantheta}
\tan2\theta=\frac{\lambda_{HS} \,\upsilon_{S}\,\upsilon_{H}}{\lambda_S\upsilon_{S}^2-\lambda_H\upsilon_{H}^2}
\eea
which is proportional to $\lambda_{HS}$ -- the strength of the SM-BSM coupling.

\section{One-Loop Corrections and Model Space}
\begin{figure}[h!]
\center
  \includegraphics[width=9cm]{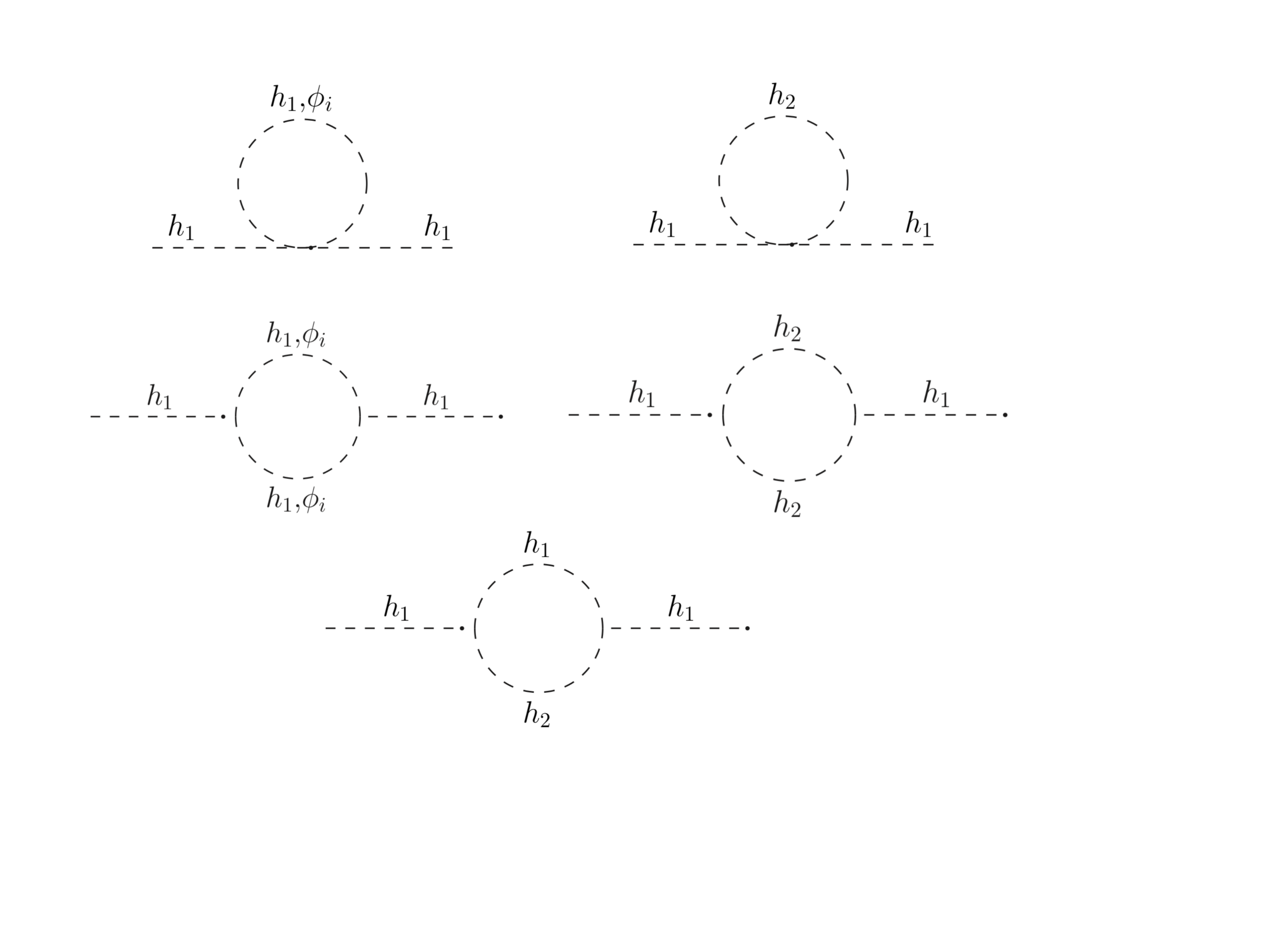}
  \caption{The one-loop diagrams leading to the $m^2_{h_1}$ corrections in (\ref{higgs0}).}\label{fig:0}
\end{figure}

In this section, we give a detailed analysis of the logarithmic corrections mentioned in (\ref{higgs0}). The Feynman diagrams which contributes the logarithmic corrections are depicted in Fig.\ref{fig:0}.
Leaving aside the quadratic corrections $c_h \Lambda^2$ in view of the symmergence mechanism mentioned in the Introduction, we keep only the logarithmic corrections ($\Lambda\gg m_{h_2}\gg m_{h_1}$)
\begin{multline}
\label{eq:logcontr}
(\delta m_{h_1}^2)_{log}=\frac{1}{8\pi^2}\left[(6\lambda_{h_1h_1h_1h_1}+3\lambda_{h_1h_1\phi\phi})m_{h_1}^2\right.\\
\hspace{3cm} \left.+(9\lambda_{h_1h_1h_1}^2
+3\lambda_{h_1\phi\phi}^2)\right]\log\left(\frac{m_{h_1}^2}{\Lambda^2}\right)\\
\hspace{1cm}+\frac{1}{16\pi^2}\left[2\lambda_{h_1h_1h_2h_2}m_{h_2}^2 + 2\lambda_{h_1h_2h_2}^2\right.\\
\left.+ \lambda_{h_1h_1h_2}^2\right] \log\left(\frac{m_{h_2}^2}{\Lambda^2}\right)
\end{multline}
where the various couplings (like the quartic couplings 
$\lambda_{h_ih_jh_kh_l}$) are listed explicitly in the Appendix 
as functions of $\lambda_H$, $\lambda_S$, $\lambda_{HS}$ and the mixing angle $\theta$. 

The $h_1$ mass receives non-trivial corrections from the $h_2$ loops. This feature, explicated in (\ref{eq:logcontr}), requires $\lambda_{HS}$ to be bounded appropriately. The vacuum stability already gives a bound (as follows from (\ref{conds})
\bea
\lambda_{HS}^2 \leq 16\lambda_H\lambda_S
\label{bound1}
\eea 
which means that $|\lambda_{HS}|$ is typically at the $30\%$ level depending on precise values of $\lambda_H$ and $\lambda_S$. We will consider different parameter ranges during the analysis. 

The bound above is however not sufficient to ensure electroweak stability. The reason is that $h_2$ can be too heavy to keep $h_1$ mass within the LHC bound. To this end, one comes back to the see-sawic bound in (\ref{seesawic0}). In what follows thus we require thus $\lambda_{HS}$ to have the value 
\bea
\label{lamhs}
\lambda_{HS} = \frac{m_{H}^2}{m_S^2}
\eea 
after expressing
\begin{multline}
m_H^2=\frac{1}{4\upsilon_H^2}\left(2\lambda_S\upsilon_S^4-4\lambda_H\upsilon_H^4-\upsilon_S^4\right.\nonumber\\
\hspace{2cm}\left.+\sqrt{8\lambda_H\upsilon_H^4\upsilon_S^4+\upsilon_S^8-4\lambda_S\upsilon_S^8+4\lambda_S^2\upsilon_S^8}\right),
\end{multline}
\begin{multline}
m_S^2= \frac{1}{4}\left(-(1+2\lambda_S)\upsilon_S^2\right.\\
\left.+\sqrt{8\lambda_H\upsilon_H^4+\upsilon_S^4-4\lambda_S\upsilon_S^4+4\lambda_S^2\upsilon_S^4} \right).
\end{multline}
as functions of the $H$ and $S$ VEVs. Trading two model parameters for the VEVs in this form leads us to the physical shell set by the VEVs. In fact, we hereon specialize to the LHC values
\begin{eqnarray}
\lambda_{H} = 0.13\;,\;\;\; \upsilon_H = 246.2\ {\rm GeV}
\end{eqnarray}
and analyze the model in terms of the remaining two free parameters: the $S$ quartic coupling $\lambda_S$ and the $S$ VEV $\upsilon_S$.

\begin{figure}[h]
  \includegraphics[width=9cm]{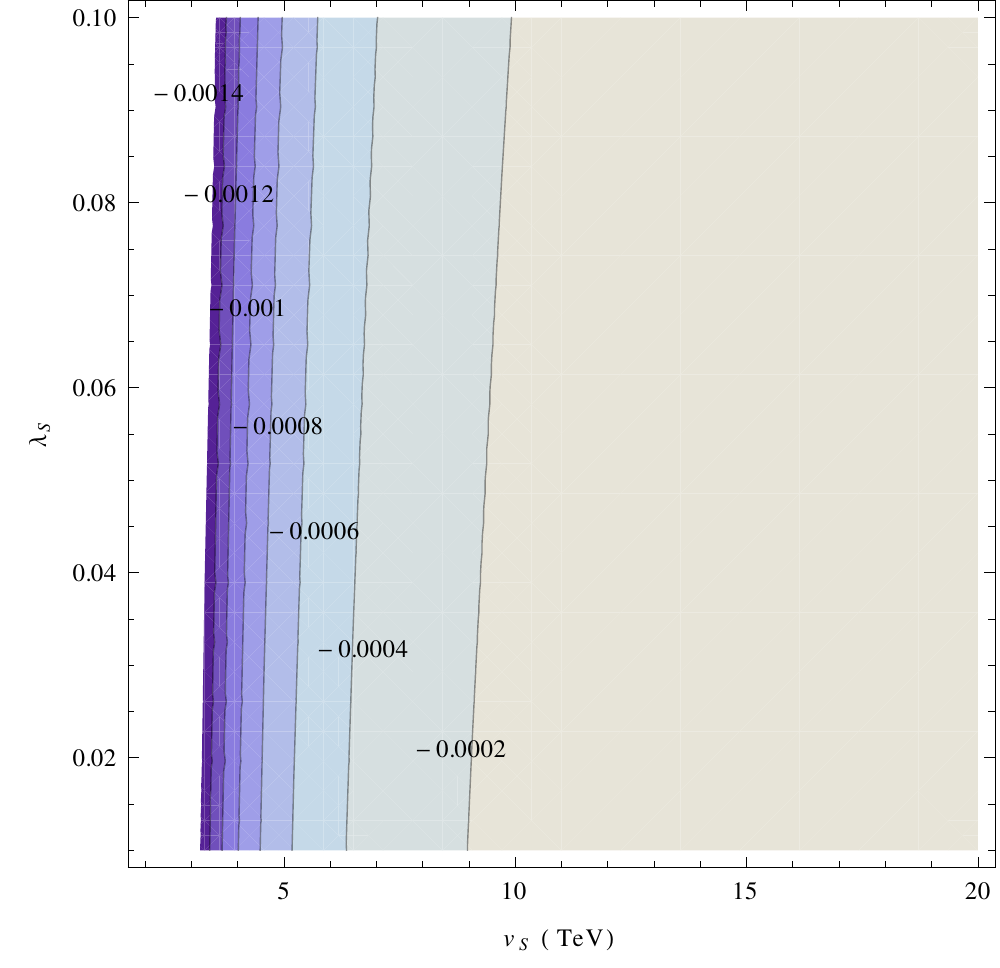}
  \caption{Variation of $\lambda_{HS}$ with $\upsilon_S\ ({\rm TeV})$ and $\lambda_S$ for  $0.01\leq \lambda_S \leq0.1$.}\label{fig:001}
\label{fig:range1}
\end{figure}

The allowed ranges of the model parameters can be determined numerically. In doing so we consider $\upsilon_S$ values as large as 20 TeV in view of the sensitivity of the exotica searches at the LHC \cite{exotica}. To this end, we plot in Fig. \ref{fig:range1} variation of $\lambda_{HS}$ with $\upsilon_S$ in the small 
$\lambda_S$ regime of  $0.01\leq \lambda_S \leq0.1$. It is seen that $\lambda_{HS}$, which decreases with $m_S^2$ due to its see-sawic  structure in (\ref{lamhs}), in magnitude, remains below $\lambda_S$ at least by two orders of magnitude. 

\begin{figure}[h]
  \includegraphics[width=9cm]{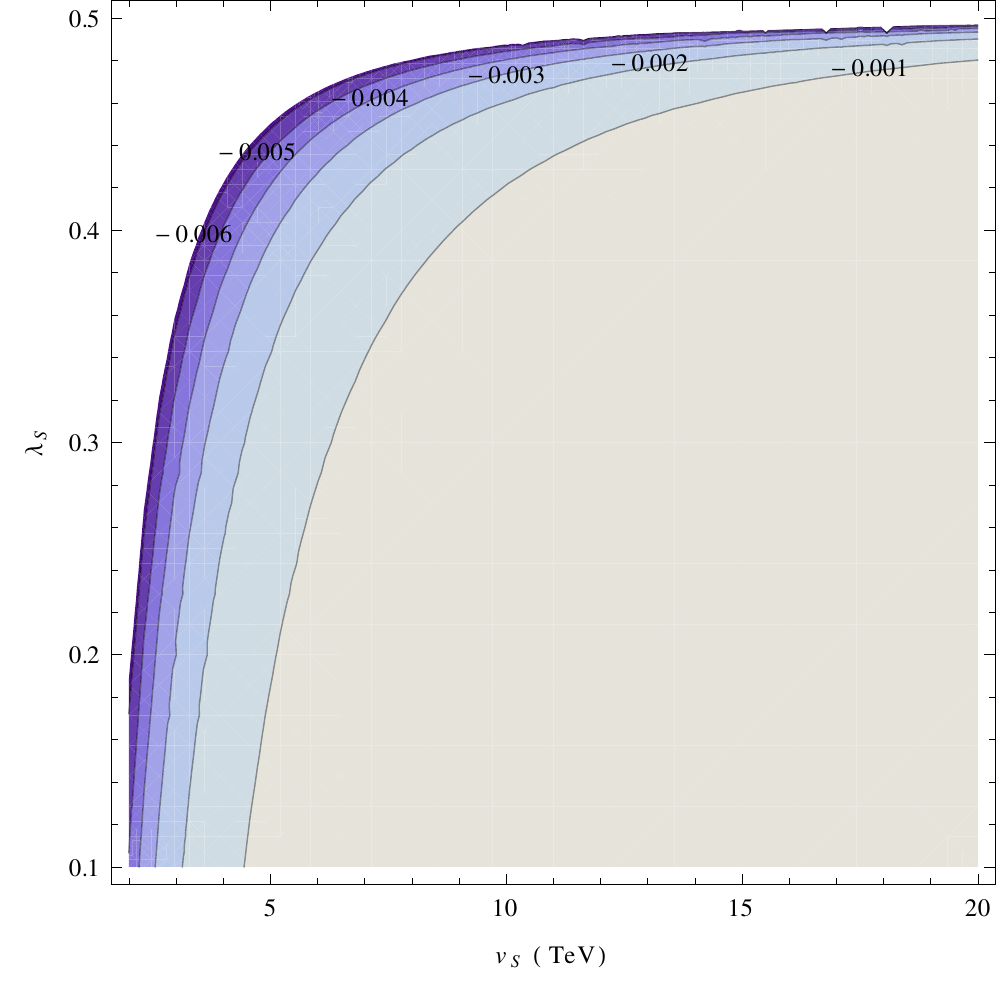}
  \caption{Variation of $\lambda_{HS}$ with $\upsilon_S\ ({\rm TeV})$ and $\lambda_S$ for  $0.1\leq \lambda_S \leq0.5$.}\label{fig:002}
\label{fig:range2}
\end{figure}

Shown in Fig. \ref{fig:range2} is the variation of $\lambda_{HS}$ with $\upsilon_S$ in the large $\lambda_S$ regime of  $0.1\leq \lambda_S \leq0.5$. It is clear that $\lambda_{HS}$, in magnitude, remains below $\lambda_S$ at least by two orders of magnitude. 

For larger $\lambda_S$, from 0.5 to 0.9, we find that $\lambda_{HS}$ takes unacceptably large values (a thousand), we do not consider therefore $\lambda_S$ values above $0.5$. In fact, hereon we set $\lambda_S=0.1$ as a nominal value revealing the physics implications of the heavy scalar. 

\begin{figure}[htbp!]
  \centering
  \includegraphics[width=9cm]{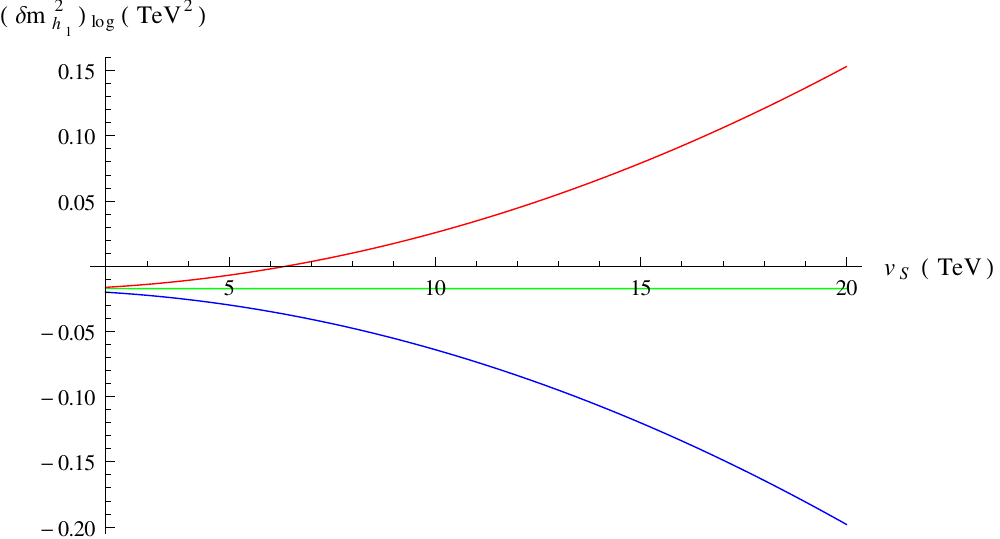}
  \caption{Corrections to the Higgs mass as a function of $\upsilon_S$ for $\lambda_{HS}=-0.01$ (red), $\lambda_{HS}=0.01$ (blue) and $\lambda_{HS}=m_H^2/m_S^2$ (green)}\label{fig:1}
\end{figure}

To see the difference between setting $\lambda_{HS}$ to a fixed (albeit small) value as in most phenomenological analyses \cite{non-inert} and requiring $\lambda_{HS}$ to obey the see-sawic bound in (\ref{lamhs}) we plot in Fig.(\ref{fig:1}) $\delta m^2_{h_1}$ in TeV as a function of $v_S$. It is clear that the see-sawic structure provides us with a rather stable electroweak scale. 

\begin{table}[htbp!]
\caption{The changes in the parameter $\lambda_{HS}$ as $\upsilon_S$ ($m_{h_2}$) increases.}
\begin{tabularx}{0.45\textwidth} { 
  | >{\centering\arraybackslash}X 
  | >{\centering\arraybackslash}X 
  | >{\centering\arraybackslash}X | }
 \hline
 $\upsilon_S$ (GeV)  & $m_{h_2}$ (GeV) & 
$\lambda_{HS}$ \\
\hline
 2000  & 894.428  & $-4.9\times 10^{-3}$ \\ 
\hline
 3000  & 1341.64  & $-2.2\times10^{-3}$  \\
\hline
 4000  & 1788.85  & $-1.2\times10^{-3}$ \\
\hline
 5000  & 2236.07  & $-8.0\times 10^{-4}$  \\
\hline
 6000  & 2683.28  & $-5.5\times 10^{-4}$ \\
\hline
 7000  & 3130.5  & $-4.0 \times10^{-4}$  \\
\hline
 8000  & 3577.71  & $-3.0\times 10^{-4}$  \\
\hline
 9000  & 4024.92  &$-2.4\times 10^{-4}$ \\
\hline
 10000  & 4472.14  & $-2.0 \times 10^{-4}$ \\
\hline
 15000  & 6708.2  & $-8.7\times10^{-5}$  \\
\hline
 20000  & 8944.27  & $-4.9\times10^{-5}$ \\
 \hline
\end{tabularx}
\label{Tab:1}
\end{table}

To see further how $\lambda_{HS}$ varies with $\upsilon_S$ we list in Table  \ref{Tab:1} $\lambda_{HS}$ values as $\upsilon_S$ ranges from 2 TeV to 20 TeV. In agreement with Figs. \ref{fig:range1} and \ref{fig:range2}, $\lambda_{HS}$ remains negative throughout and well satisfies the vacuum stability bound (\ref{bound1}). It is clear that larger the $m_S$ of scalar field, the weaker its interaction with Higgs. This decrease could explain why we have not observed any fingerprint of BSM physics (the scalar $S$ here) at LHC experiments.    \\

Before closing this section, it is worth noting that any value of $\lambda_{HS}$ obeying the constraint   (\ref{seesawic0}) can satisfy the electroweak stability. However, it is not difficult to see that smaller $\lambda_{HS}$ ($<m_H^2/m_S^2$) leads to the requirement of larger luminosities and energies for the discovery of new physics at colliders. The maximal value of $\lambda_{HS}$ ($=m_H^2/m_S^2$) provides one with accessible and realistic luminosities and energies for discovery. This is the reason why we have considered $\lambda_{HS}$ in (\ref{lamhs}) throughout our analyses.

\section{Collider Phenomenology}\label{analysis}
In this section we perform a detailed simulation study to answer the question  (\ref{question}) in the Introduction. The analysis involves production and decay rates as well as event selection and background analysis. Below is a systematic discussion of the analysis stages.

\begin{figure}[htbp]
\begin{center}
\includegraphics[scale=.40]{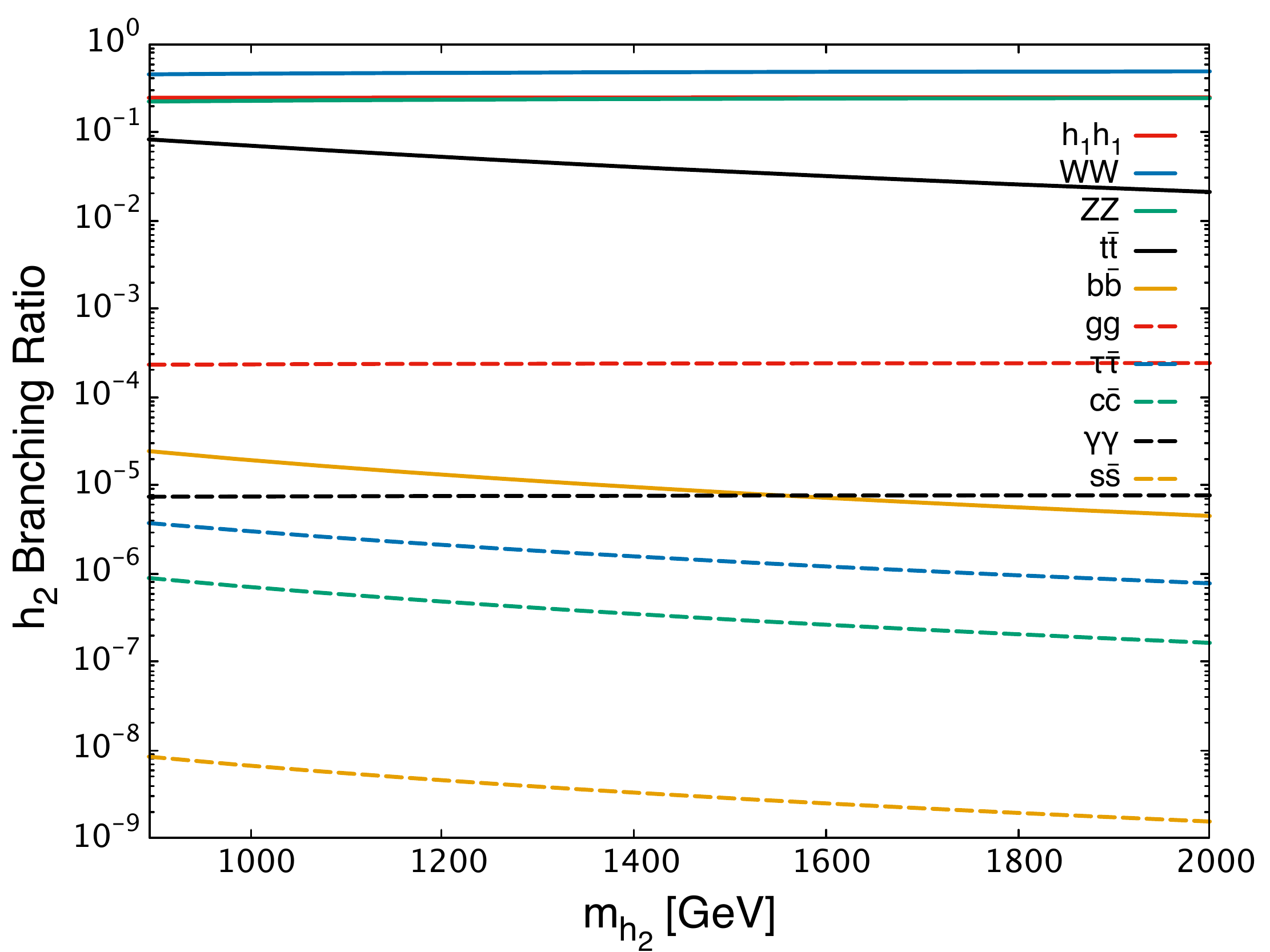}
\caption{Branchings of the heavy scalar $h_2$ into various SM particles. The $h_{1}h_{1}$, WW and ZZ are the dominant decay modes. 
The decay rates remain constant essentially or decrease with $m_{h_{2}}$ due to the see-sawic coupling.}
\label{fig:br}
\end{center}
\end{figure}

The production cross section of real singlet scalar depends on its mass and its
coupling to the SM Higgs field. In view of the see-sawic coupling (\ref{lamhs})
the production cross section is directly set by $m_{h_2}$ (or $v_S$). It sets also
branching fractions of $h_2$ decays. The braching fractions of $h_2$ into various SM particles are given in Fig. \ref{fig:br}. The dominant decay channels are seen to be $h_2 \rightarrow WW$ ($49\%$), $h_2 \rightarrow h_1 h_1$ ($25\%$)
and $h_2 \rightarrow Z Z$ ($24\%$), which are almost independent of $m_{h_2}$.
This constancy of the branching fractions, a property following from the seesawic couplings in (\ref{seesawic2}) and (\ref{linkup}), proves useful for putting discovery limits
(as in the simplified models \cite{simplified}).

\begin{figure}[htbp]
\begin{center}
\includegraphics[scale=.40]{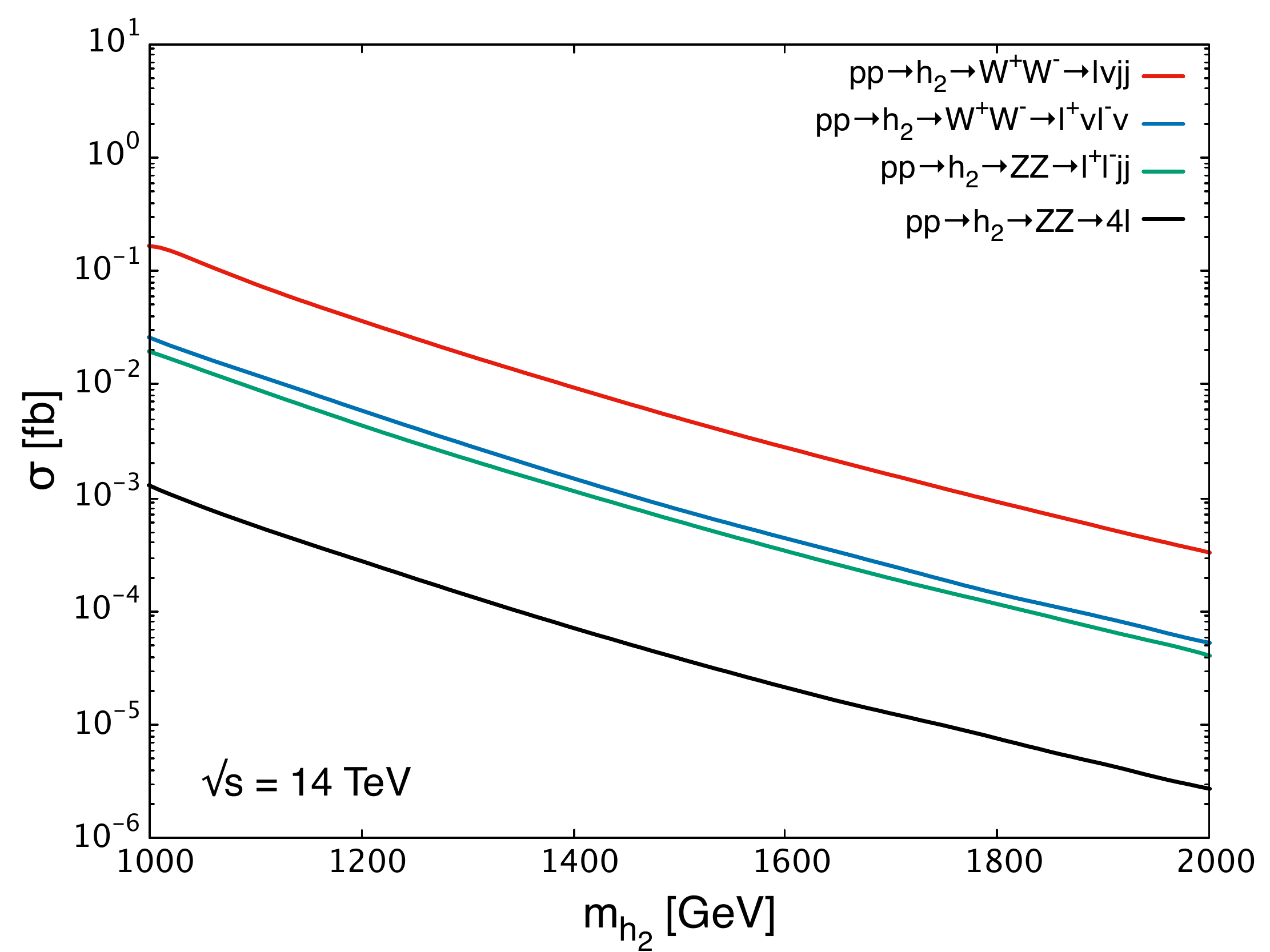} \\
\includegraphics[scale=.40]{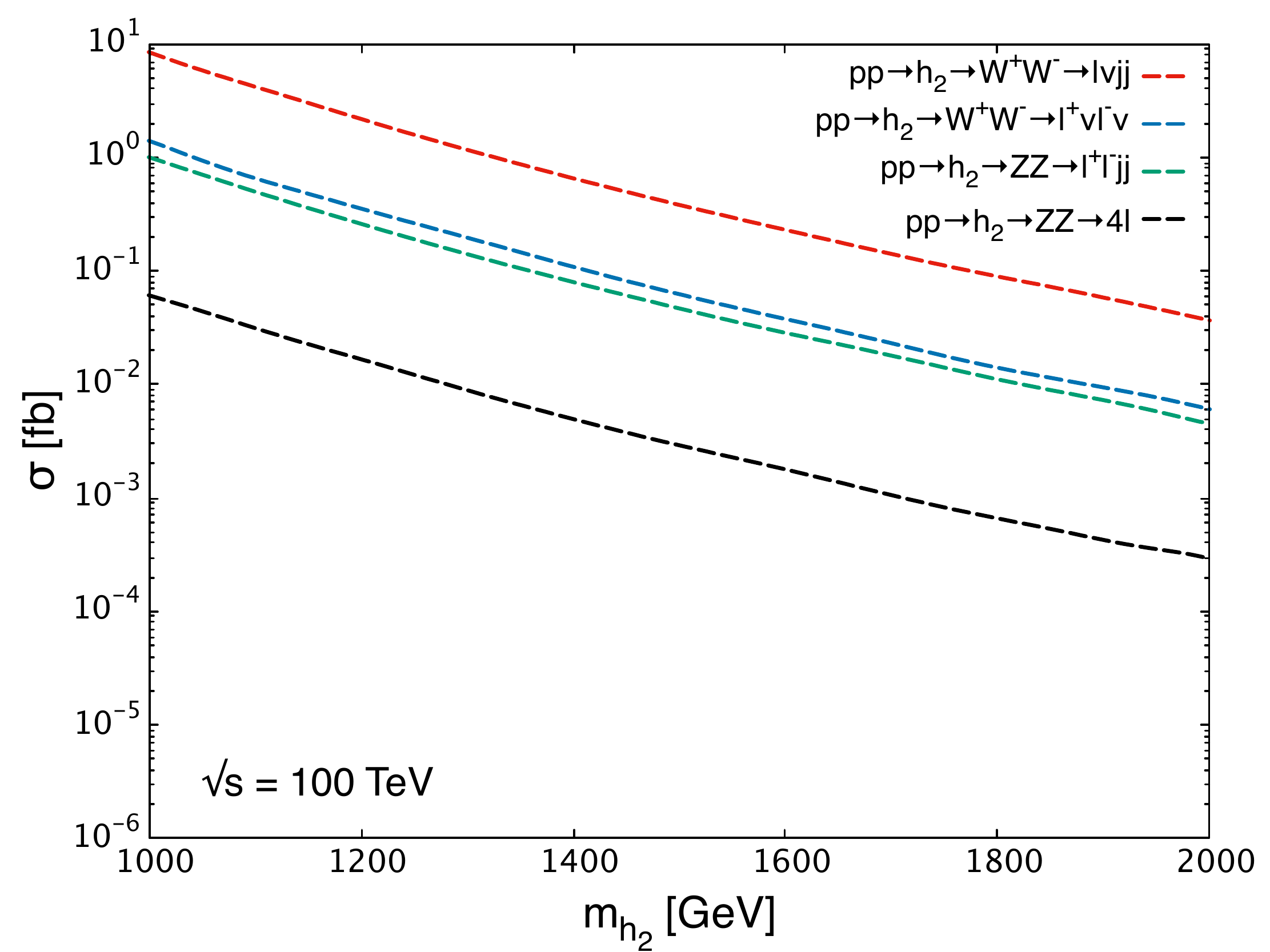}
\caption{Cross section (fb) times branching ratio as functions of $m_{h_{2}}$ for the channels $pp \to h_2 \to W^+W^- \to l\nu j j$ (red line), $pp \to h_2 \to W^+W^- \to l^+ \nu l^- \nu$ (blue line), $pp \to h_2 \to ZZ \to l^+ l^- j j$ (green line) and $pp \to h_2 \to ZZ \to 4l$ (black line) for $\sqrt{s}=14$~TeV (up) and $\sqrt{s}=100$~TeV (down).}
\label{fig:cs}
\end{center}
\end{figure}

In Fig.~\ref{fig:cs}, $h_2$ production cross sections times branching ratios are given 
as a function of $m_{h_{2}}$ for the $pp \to h_2 \to W^+W^- \to l\nu j j$, 
$pp \to h_2 \to W^+W^- \to l^+ \nu l^- \nu$, $pp \to h_2 \to ZZ \to l^+ l^- j j$ and 
$pp \to h_2 \to ZZ \to 4l$ channels separately for $\sqrt{s}=14$~TeV and $\sqrt{s}=100$~TeV. Although $pp \to h_2 \to W^+W^- \to l\nu j j$ channel has the highest cross section, reconstruction of this type of semi-leptonic final states {\it i.e.} a charged lepton (electron or muon) is challenging due to the
large missing transverse momentum coming from the presence of a neutrino in each event 
and at least two jets.  In addition, W+jets background is dominant and gives a peak in the same region 
with diboson invariant mass, making it extremely difficult to separate signal from the background ~\cite{Aad:2012me}.
The $pp \rightarrow h_2 \rightarrow W^+W^- \rightarrow l^+ \nu l^- \nu$ channel, which has the second highest cross section, 
is another challenging channel since the invariant mass of the system is not completely 
reconstructable due to the missing energy in the final states coming from the neutrinos~\cite{Aad:2019lpq}.
 
In $pp \to h_1 h_1$ channel, it is possible to search for $h_2$ via $4\gamma$ (suppressed by loops), $\gamma\gamma b \bar{b}$ (suppressed by loops and small bottom Yukawa), and the like. We will not take into account these channels in our analysis.
 
The $pp \to h_2 \to ZZ \to l^+ l^- j j$ channel has a higher cross section than $pp \to h_2 \to ZZ \to 4l$ 
and therefore may bring  more statistics, but due to the jets in the final states, this channel has larger background especially from Z+jets and $t\bar{t}$ processes. 
The $pp \to h_2 \to ZZ \to 4l$ has smaller cross sections than the other channels presented in Fig.~\ref{fig:cs},
but it has completely reconstructable final state (four charged leptons) and has lower background contamination ~\cite{Aad:2014wra,Collaboration:2012iua}. Therefore, we focus on the $4l$ channel and perform a detailed analysis using various different mass scenarios.  

In total, we consider seven different mass scenarios between 800-2000 GeV in increments of 200 GeV,
for the search of $h_{2}$ via the $4l$ decay channel. We present kinematic distributions and the event selection 
efficiencies for the low mass and high mass scenarios, $m_{h_2}=800$ GeV and $m_{h_2}=2000$ GeV. The final results are given for every mass value considered in the analysis. 
All analyses are carried out at two different center of mass energies;
at $\sqrt{s}=14$~TeV and $\sqrt{s}=100$~TeV, corresponding, respectively, to  the center of mass energy of the $pp$ collisions
at the high luminosity phase of the LHC (HL-LHC) and the future circular collider FCC-hh \cite{Aleksa:2019pvl,Schmidt:2016jra,Atlas:2019qfx}. 

\begin{figure}
\centering
\subfloat{\includegraphics[scale=.25]{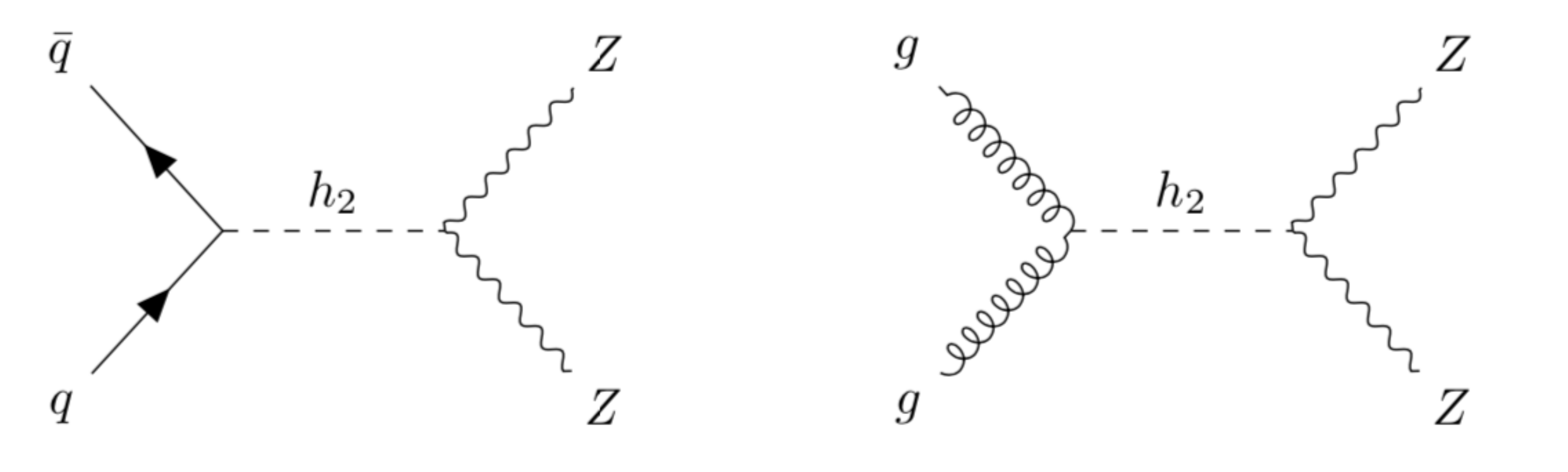}\label{fig:sig_diagram}}~\\
\subfloat{\includegraphics[scale=.25]{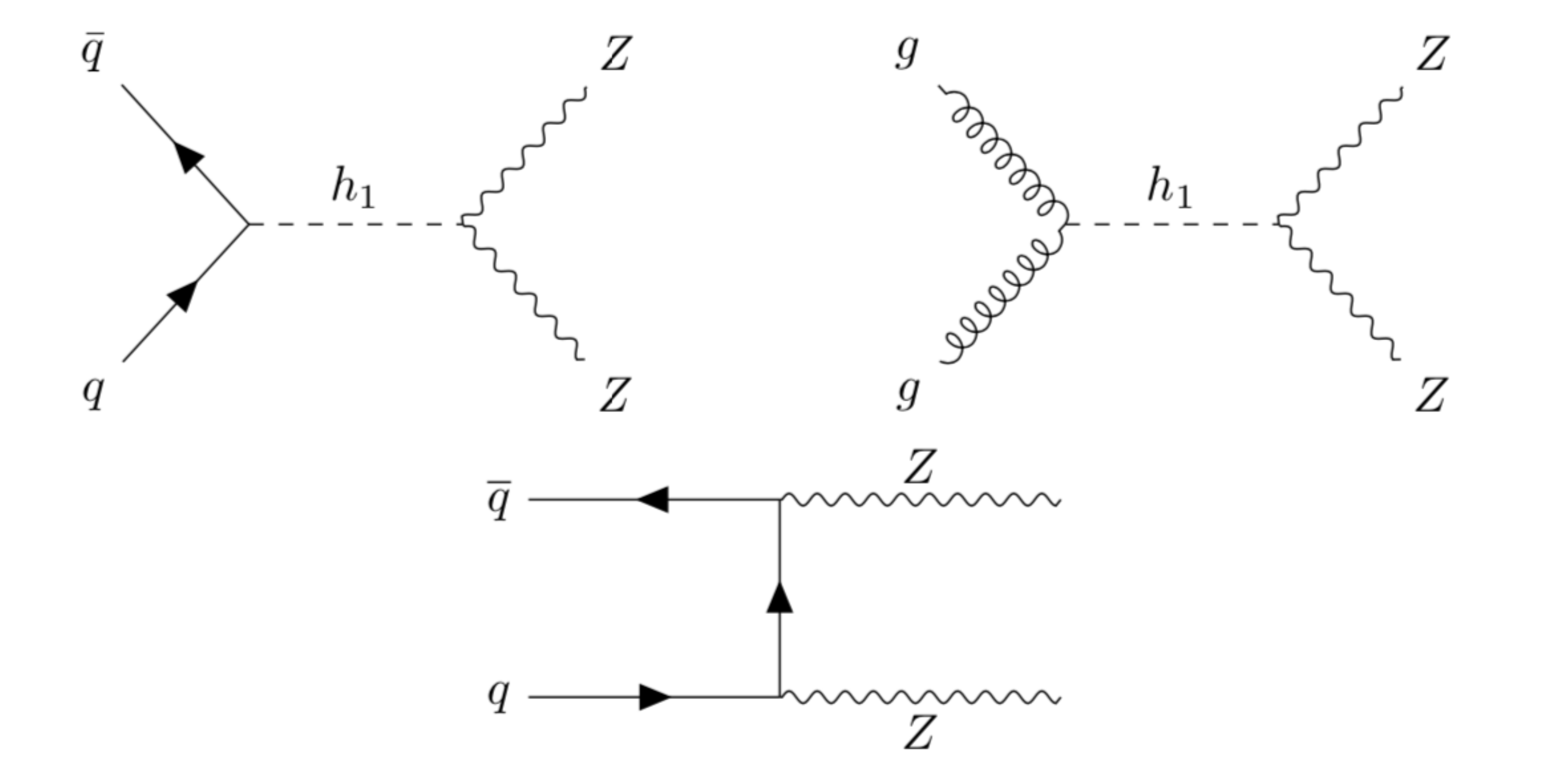}\label{fig:bkg_diagram}}
\caption{The leading order Feynman diagrams for the signal process (a) and the main background process (b).}
\label{fig:diagrams}
\end{figure}

In simulating the $4l$ signal and the SM background events for $pp \to ZZ$, whose Feynman diagrams are depicted in Fig.~\ref{fig:diagrams},
we have modified the SM package in LanHEP v3.2.0~\cite{lanhep} by including the real singlet $S$, and exported the extended model to CalcHEP v3.7.5~\cite{CalcHEP}. The events are simulated by CalcHEP using the LHAPDF v6.1.6~\cite{Buckley:2014ana} library and
its CTEQ6L1~\cite{cteq6l1} parton distribution functions (PDFs) as well as 
PYTHIA8 v2.3.0~\cite{Sjostrand:2014zea} for parton showering and hadronization.
The detector response is simulated by Delphes v3.4.2 fast-simulation package~\cite{deFavereau:2013fsa} using 
the HL-LHC and FCC-hh detector card files implemented in it. Events are analyzed using the ExRootAnalysis 
package linked to ROOT v6.12~\cite{Antcheva:2009zz}.

To search for the heavy scalar $h_2$ in $pp \to h_2 \to ZZ \to 4l$ channel, we select events 
with four leptons in the final state, where each lepton is either an electron or a muon.
The two Z bosons in the events, each decaying to 
opposite-sign, same-flavor (electrons or muons) lepton pair,  lead to three possible final states configurations; $4e$, $4\mu$, and $2e2\mu$. In each selected event, two Z boson candidates are reconstructed from lepton pairs considering all possible  pairing combinations and the one with the mass closer to the mass of the Z boson is called $\text{Z}_1$ and the other reconstructed Z boson is $\text{Z}_2$. Following a similar analysis strategy and event selection as in \cite{Sirunyan:2017zjc,Banerjee:2015hoa}, we applied the pre-selection cuts listed in Table~\ref{Tab:precuts}.
In this table, $\Delta \text{R}$ is the distance between two leptons in the $\eta$-$\phi$ plane and defined as
$\Delta \text{R}(l_i,l_j) = \sqrt{(\eta_{i}-\eta_{j})^2 - (\phi_{i}-\phi{j})^2}$ where $i$ and $j$
refer to the lepton pairs used in reconstruction of the Z bosons.  

\begin{table}
\centering
\caption{Pre-selection cuts for identifying the leptons ($l = e \text{ or }\mu$).}
 \begin{tabular}{l|l}
 \hline
	Quantity & Selection criteria \\
 \hline
	Transverse momentum  & $p_{\text{T}}^l > 10$~GeV \\
    Pseudo-rapidity      & $|\eta_{l}|<$ 2.5 \\	
    Radial distance      & $\Delta \text{R}({l_i, l_j}) > 0.2$ \\ 
 \hline
\end{tabular}
\label{Tab:precuts}
\end{table}

\begin{figure}
\centering
{\includegraphics[scale=.20]{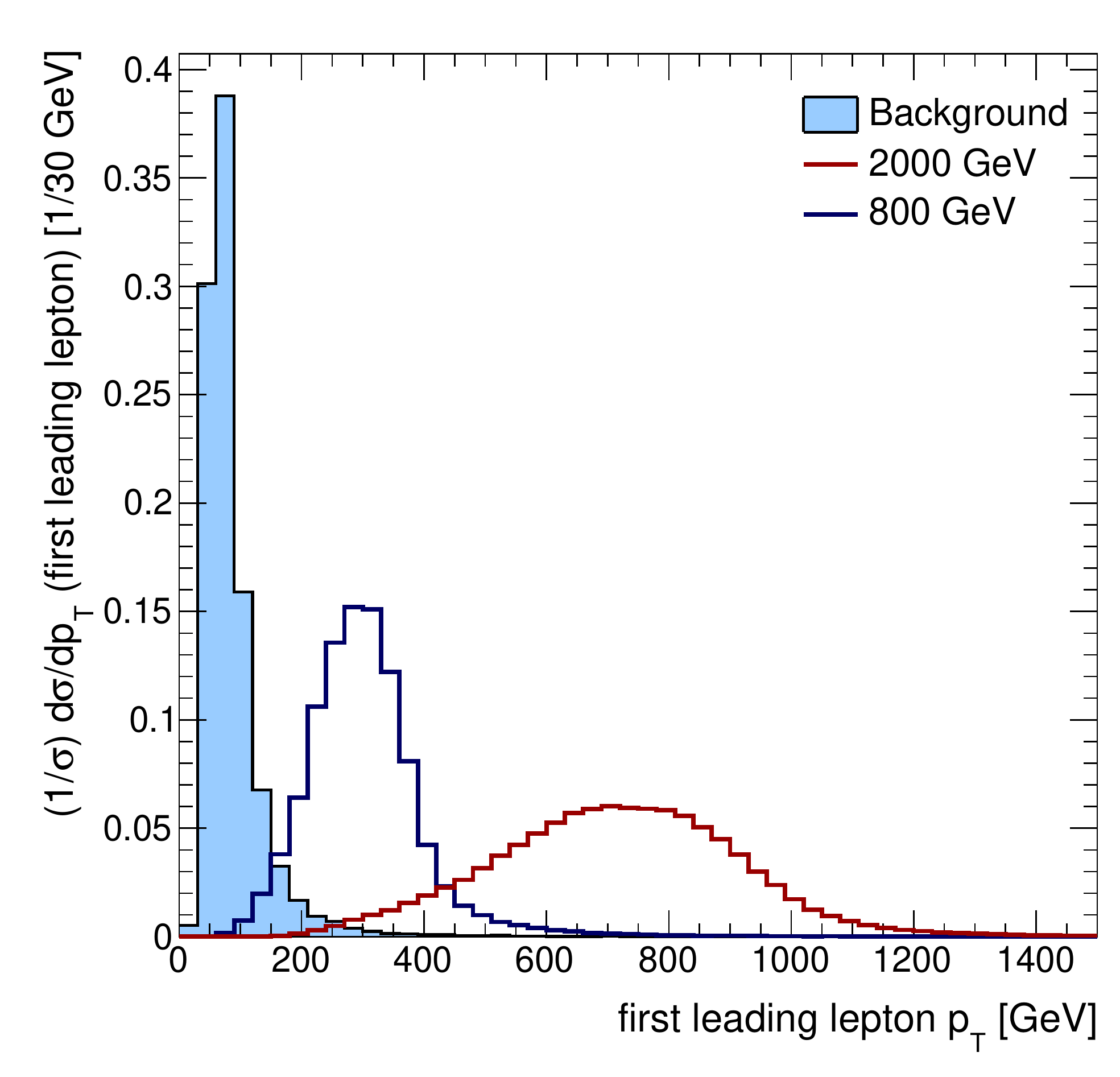}\label{fig:pTl1}}~
\subfloat{\includegraphics[scale=.20]{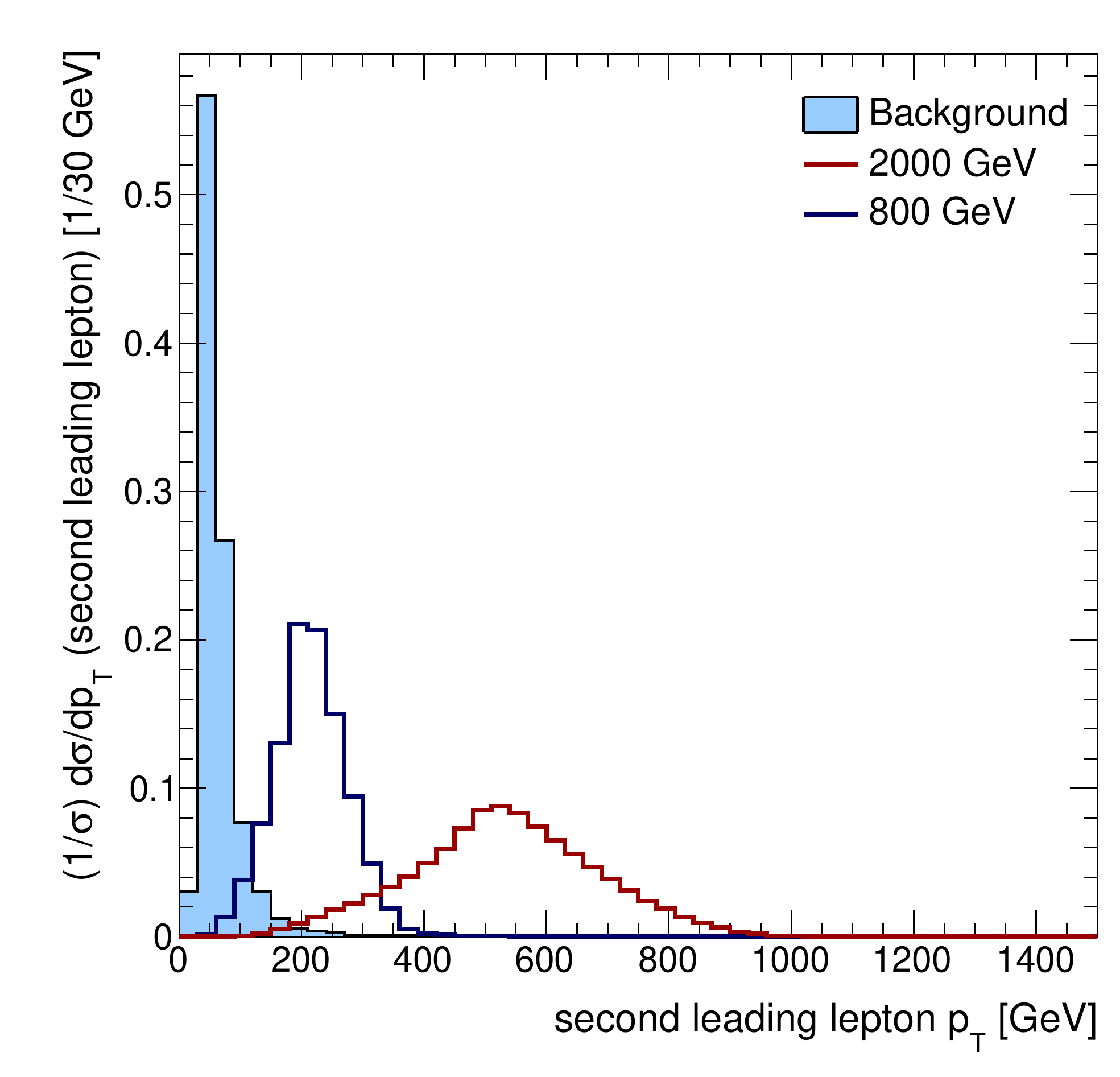}\label{fig:pTl2}}~\\
\subfloat{\includegraphics[scale=.20]{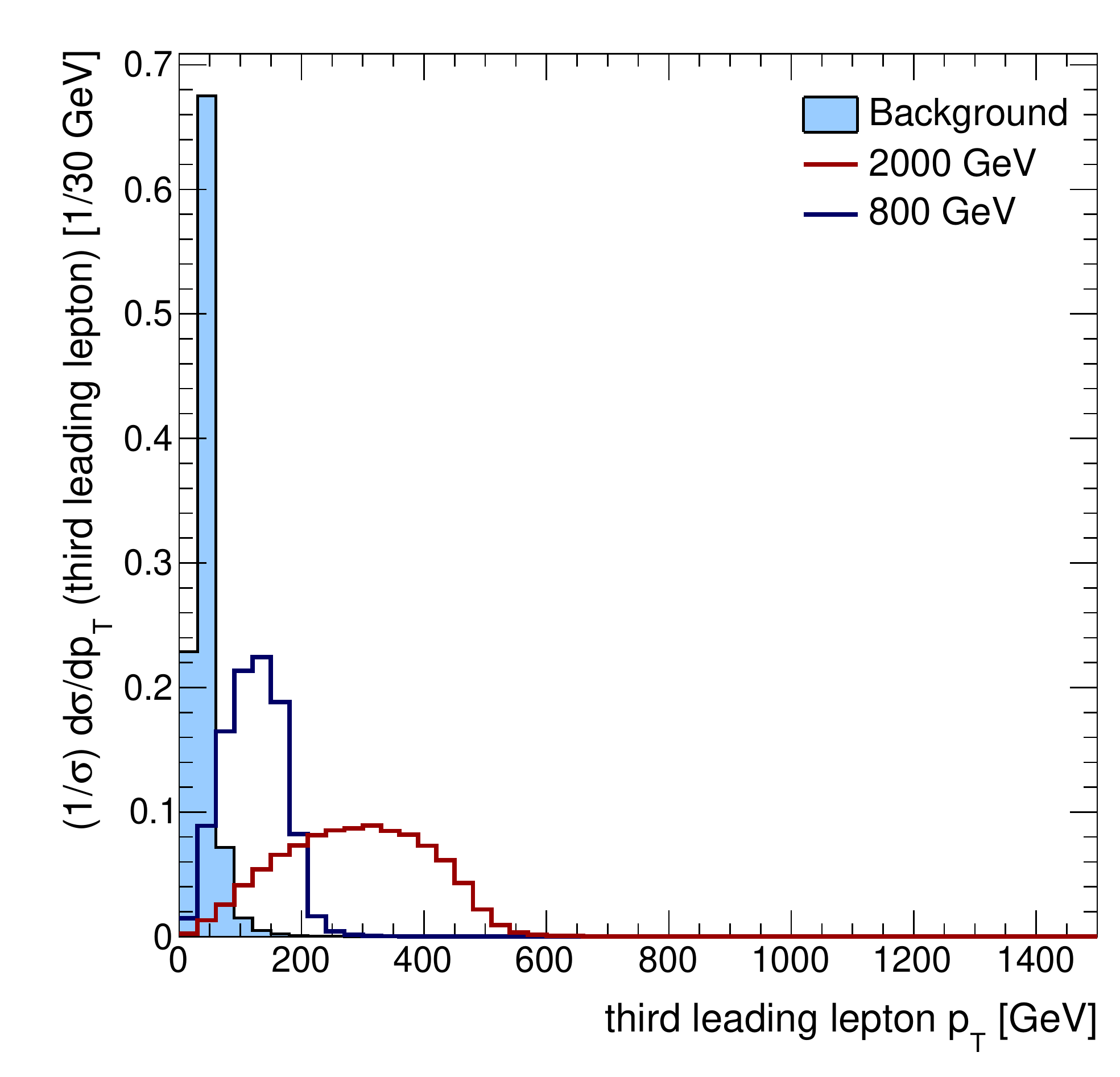}\label{fig:pTl3}}~
\subfloat{\includegraphics[scale=.20]{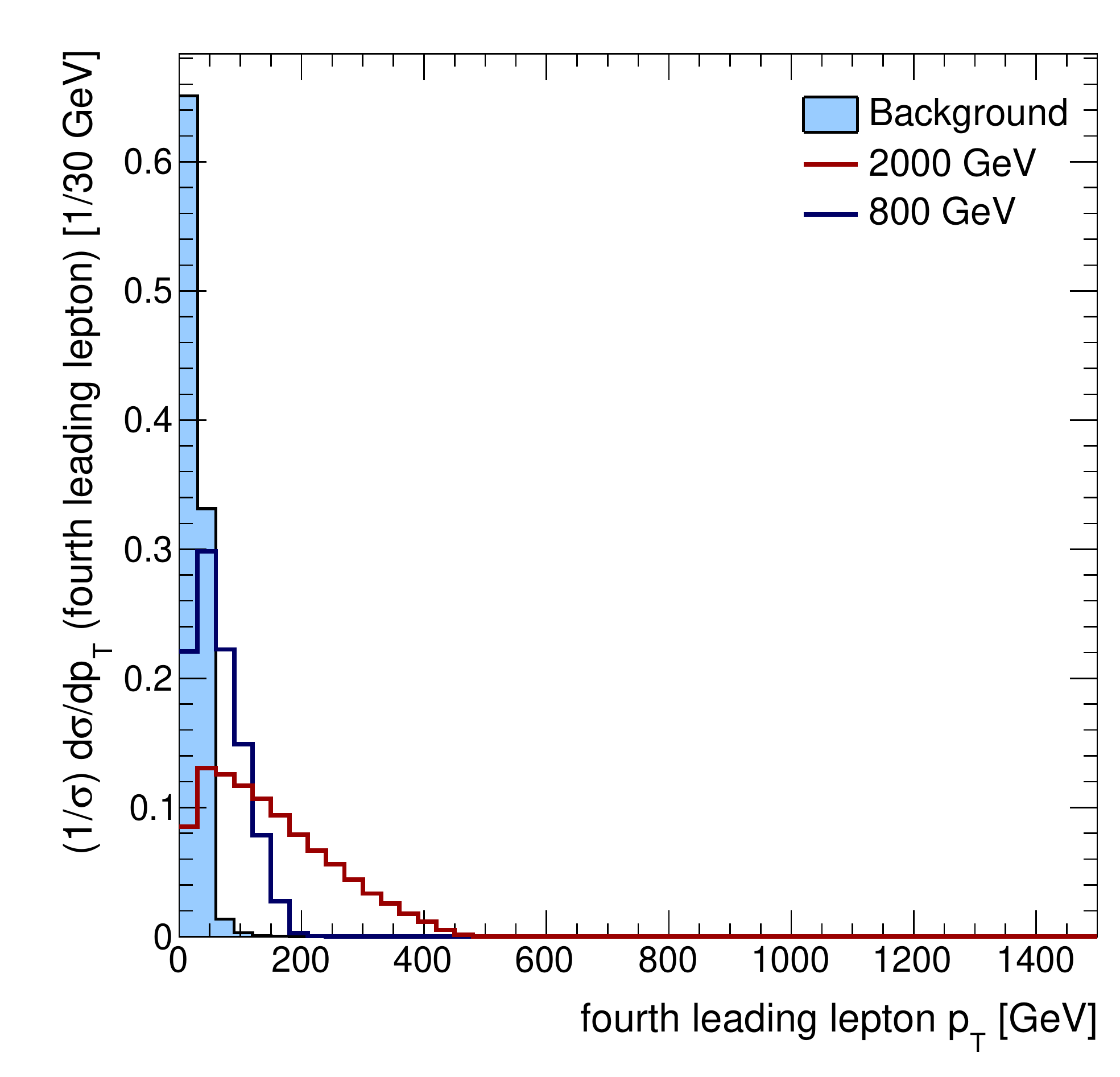}\label{fig:pTl4}}
\caption{Normalized differential cross sections as a function of leading lepton $p_{T}$ (a), 
second-leading lepton $p_{T}$ (b), third-leading lepton $p_{T}$ (c) and fourth-leading lepton $p_{T}$ (d)
for $m_{h_{2}}=2000$~GeV (red line), $800$ GeV (blue line) and background (blue solid) at $\sqrt{s}=14$~TeV.}
\label{fig:lpt14}
\end{figure}

\begin{figure}[htbp]
\centering
\subfloat{\includegraphics[scale=.20]{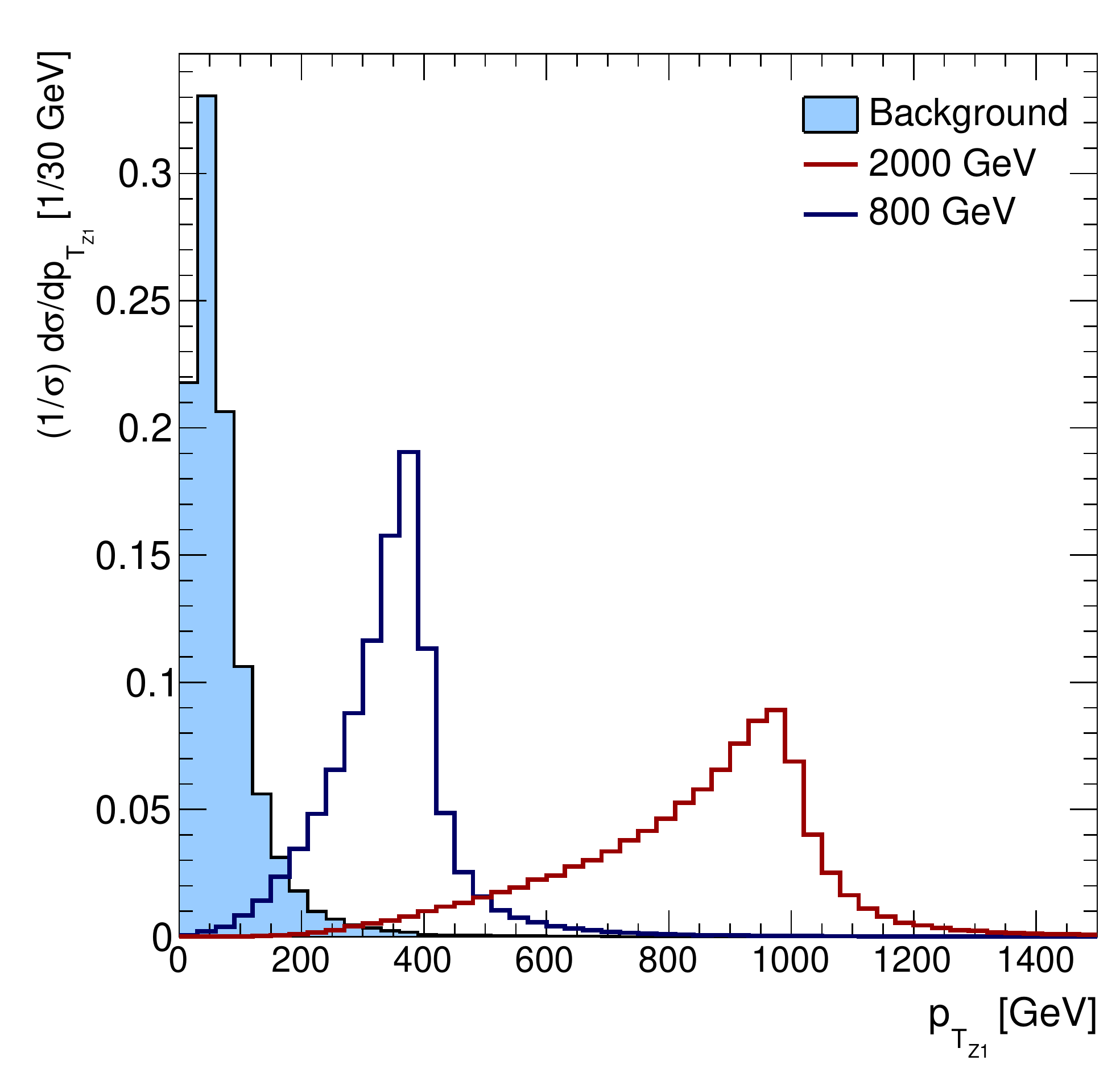}\label{fig:14TeV_Z1}}~
\subfloat{\includegraphics[scale=.20]{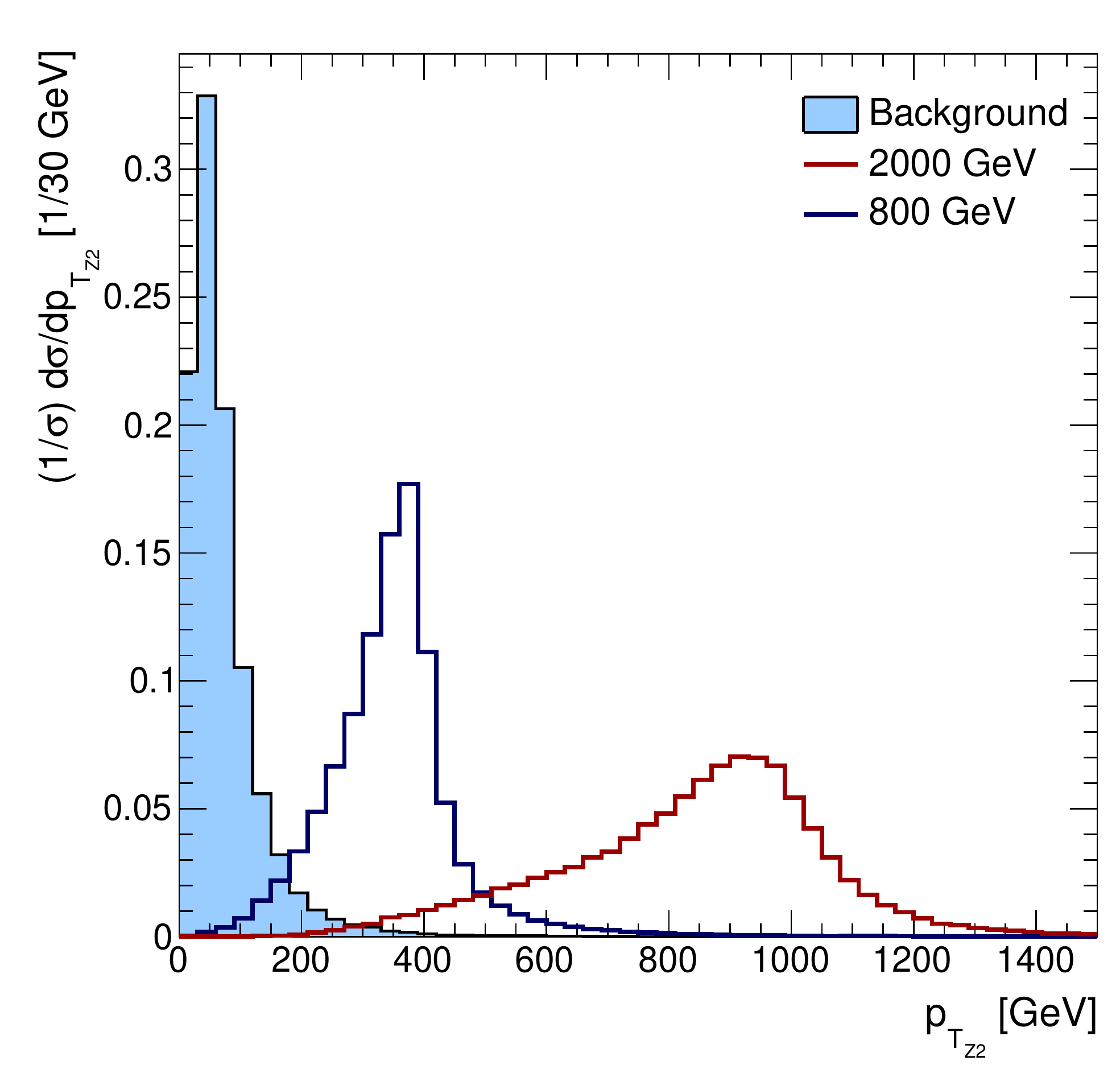}\label{fig:14TeV_Z2}}~\\
\subfloat{\includegraphics[scale=.20]{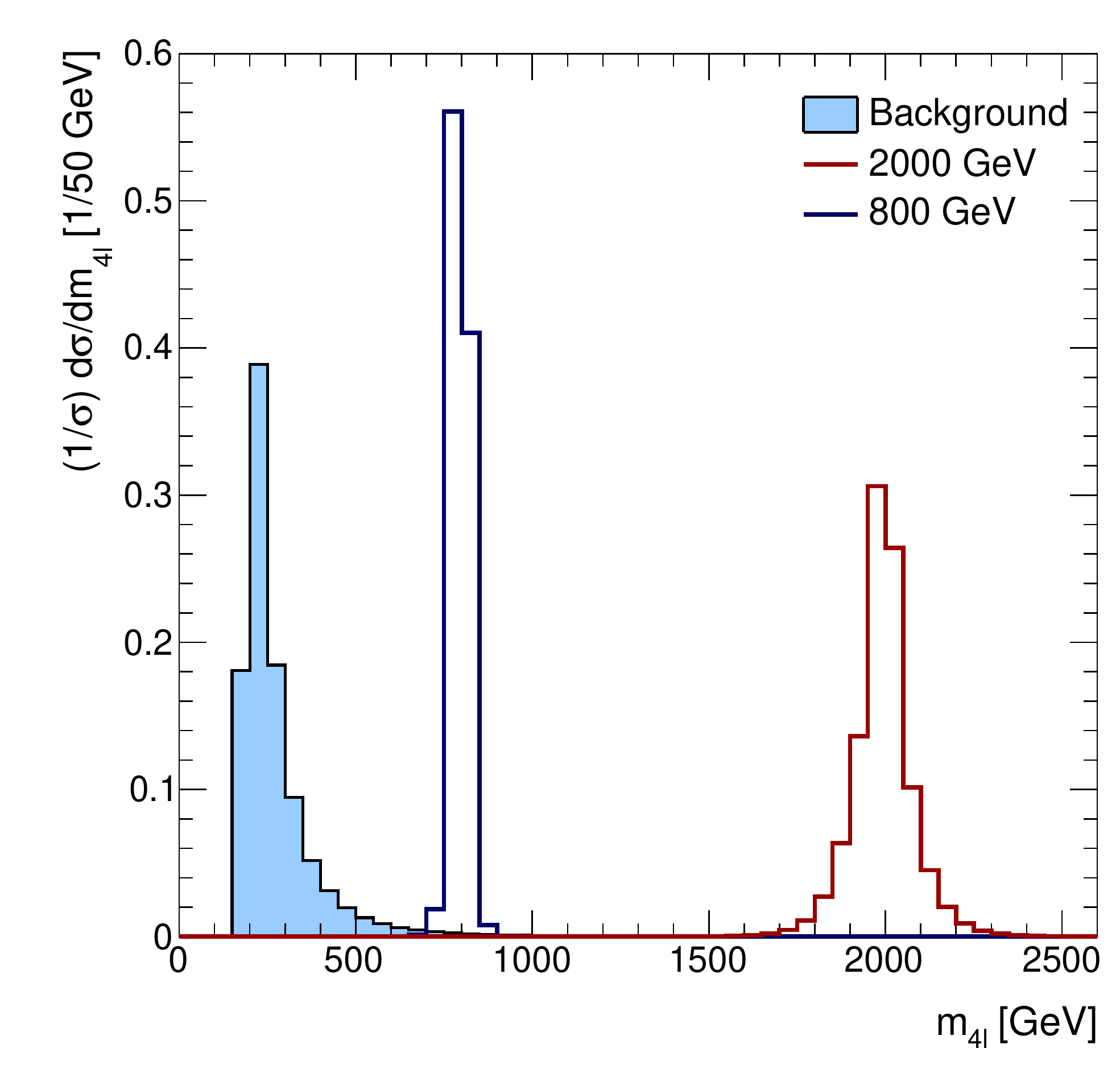}\label{fig:14TeV_m4l}}~
\subfloat{\includegraphics[scale=.20]{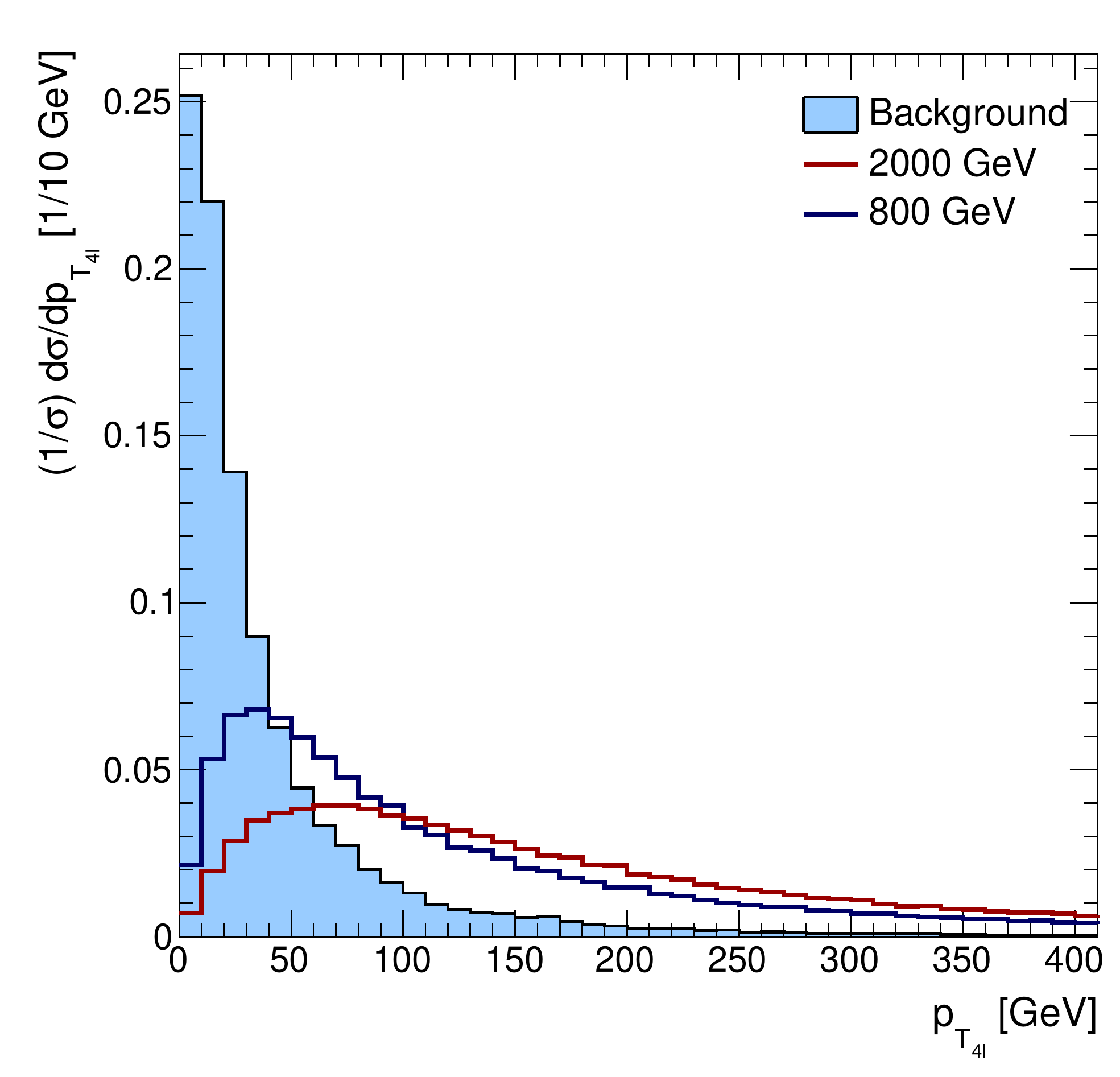}\label{fig:14TeV_pt4l}}
\caption{Normalized differential cross sections as a function of the transverse momentum of $\text{Z}_1$ boson (a) 
$\text{Z}_2$ boson (b), and the the invariant mass (c) and transverse momentum (d) 
of the four-lepton system for $m_{h_{2}}=2000$~GeV (red line), $800$ GeV (blue line) 
and background (blue solid) at $\sqrt{s}=14$~TeV.} 
\label{fig:Z14}
\end{figure}

\begin{figure}
\centering
\subfloat{\includegraphics[scale=.20]{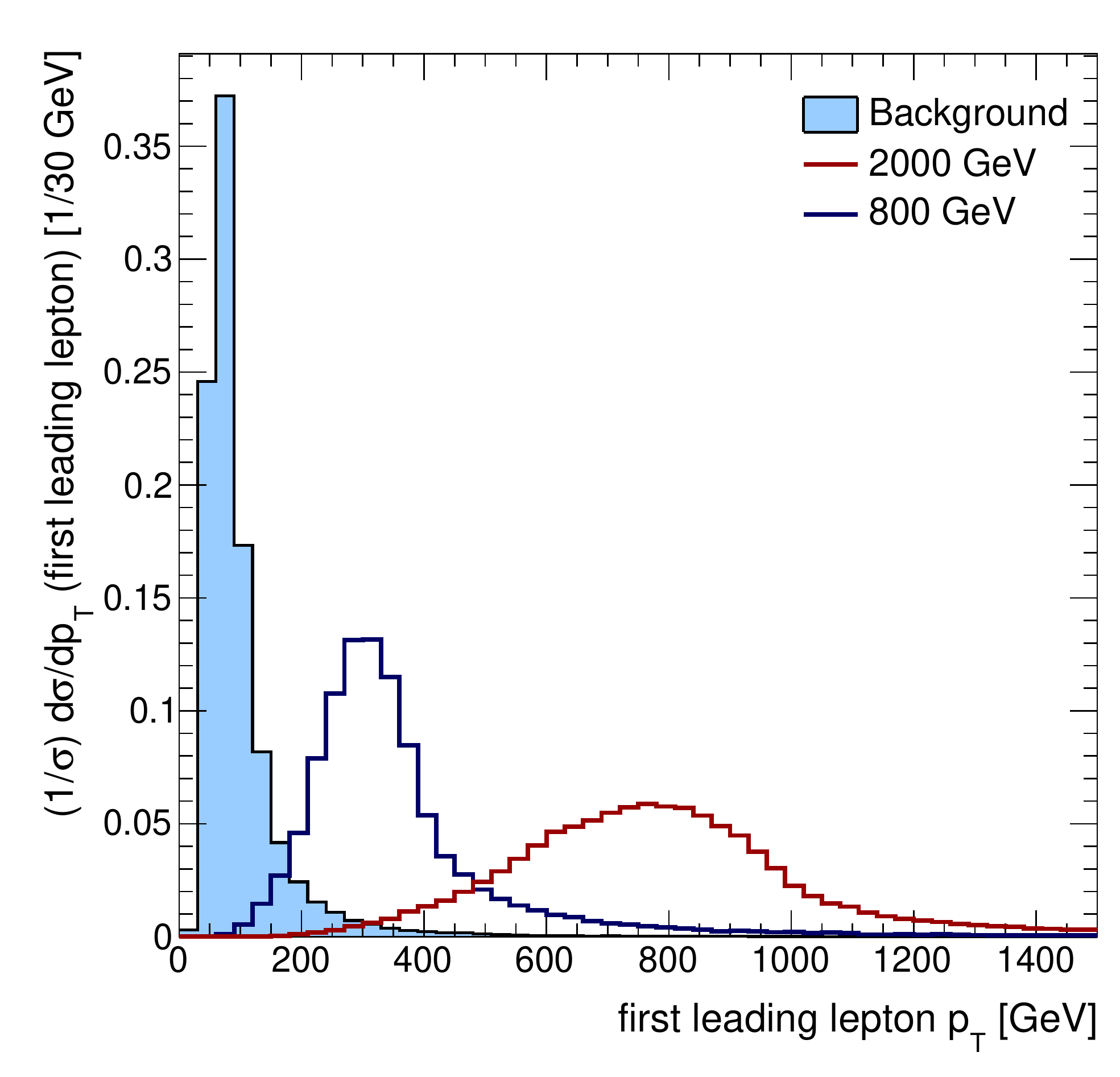}\label{fig:pTl1}}~
\subfloat{\includegraphics[scale=.20]{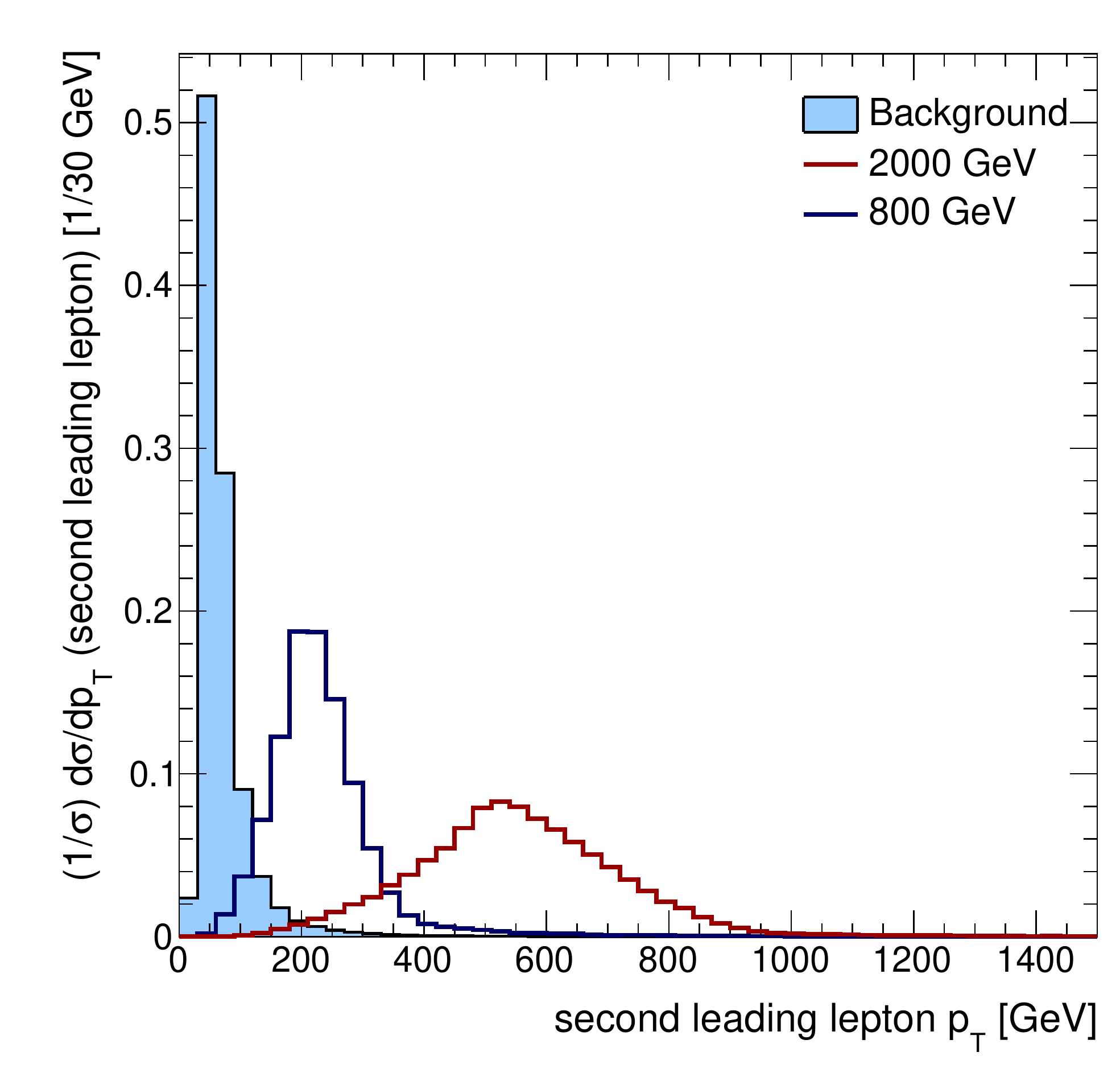}\label{fig:pTl2}}~\\
\subfloat{\includegraphics[scale=.20]{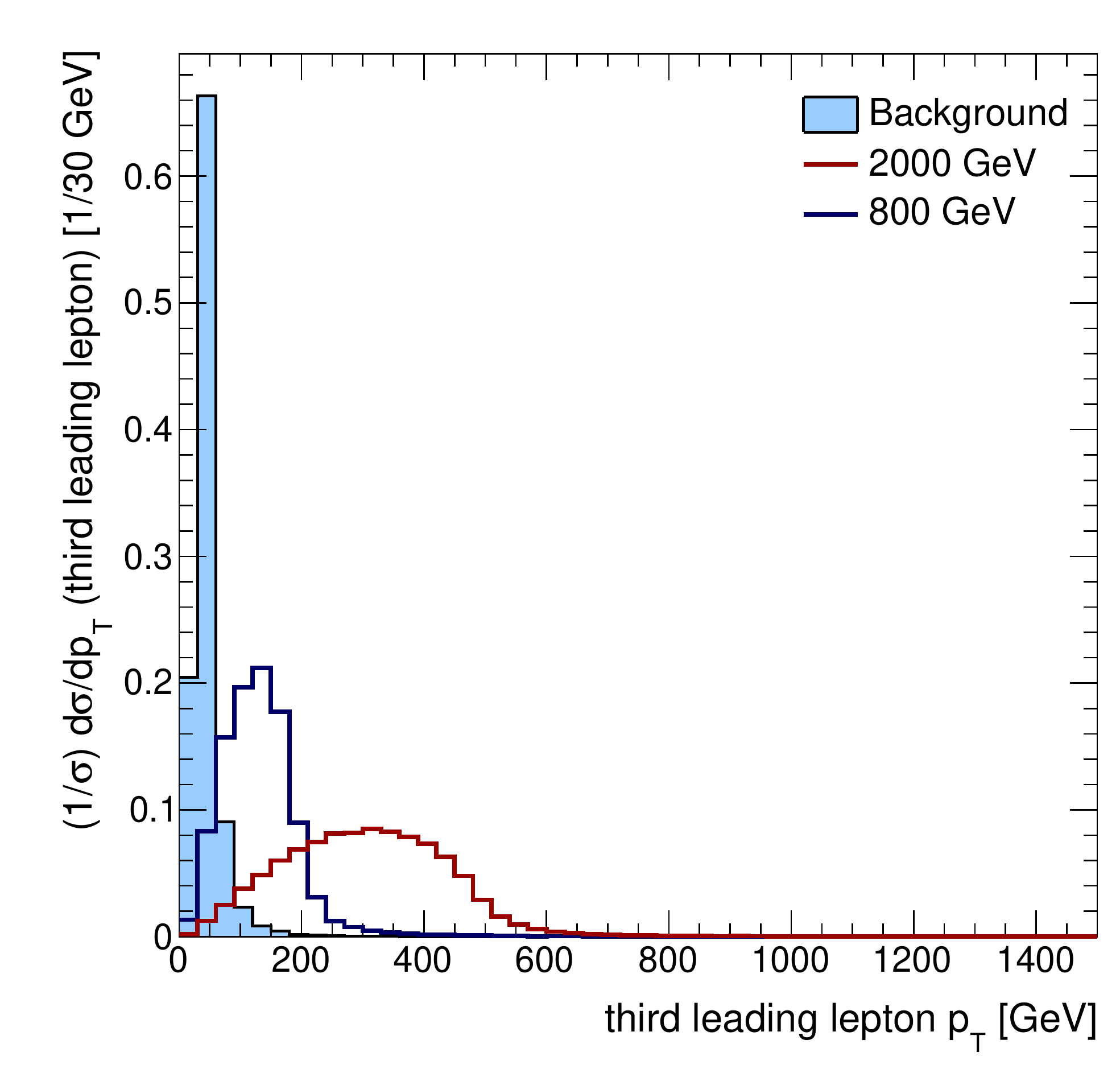}\label{fig:pTl3}}~
\subfloat{\includegraphics[scale=.20]{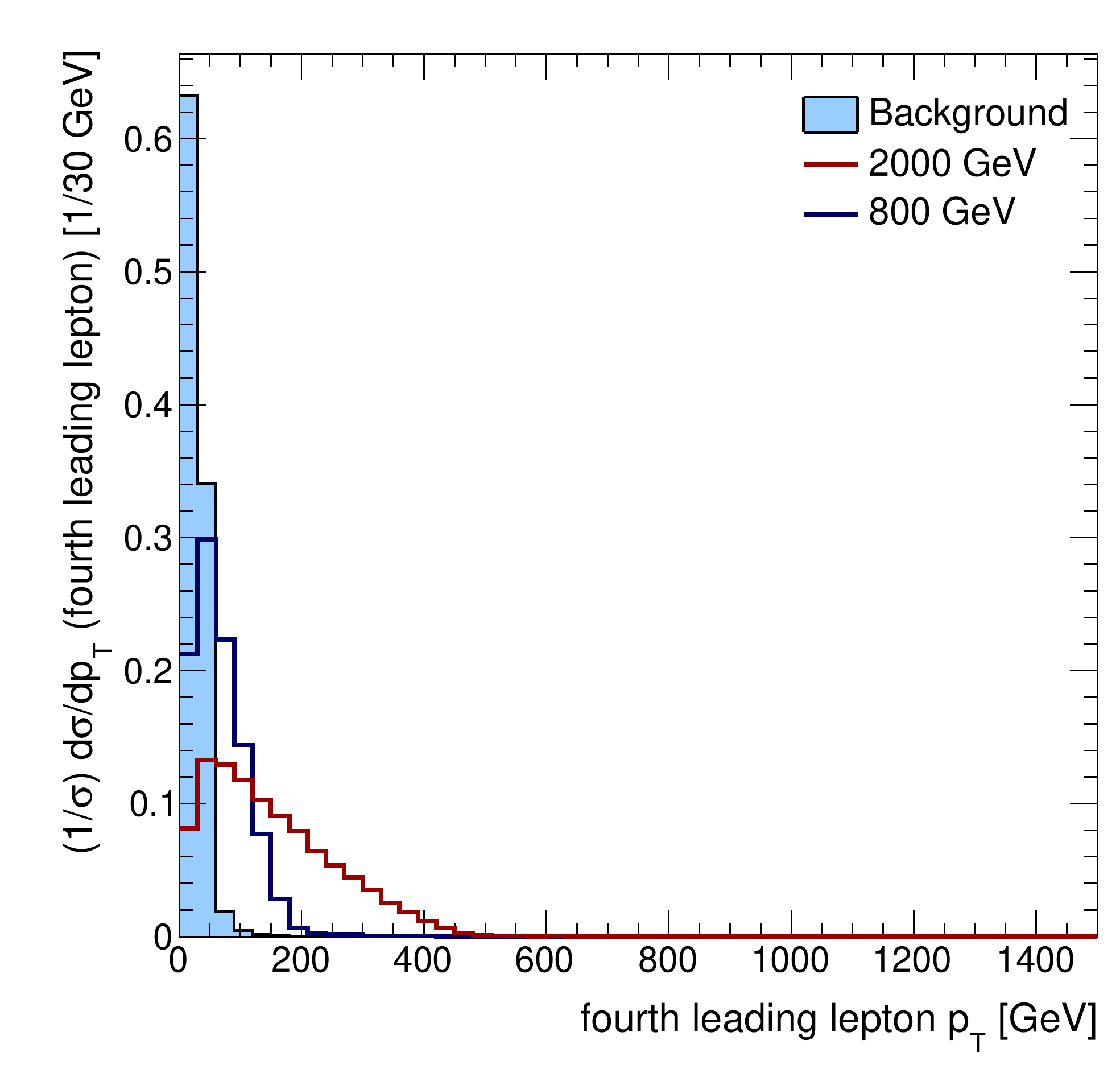}\label{fig:pTl4}}
\caption{Normalized differential cross sections as a function of leading lepton $p_{T}$ (a), 
second-leading lepton $p_{T}$ (b), third-leading lepton $p_{T}$ (c) and fourth-leading lepton $p_{T}$ (d)
for $m_{h_{2}}=2000$~GeV (red line), $800$ GeV (blue line) and background (blue solid) at $\sqrt{s}=100$~TeV.}
\label{fig:lpt100}
\end{figure}

\begin{figure}[htbp]
\begin{center}
\subfloat{\includegraphics[scale=.20]{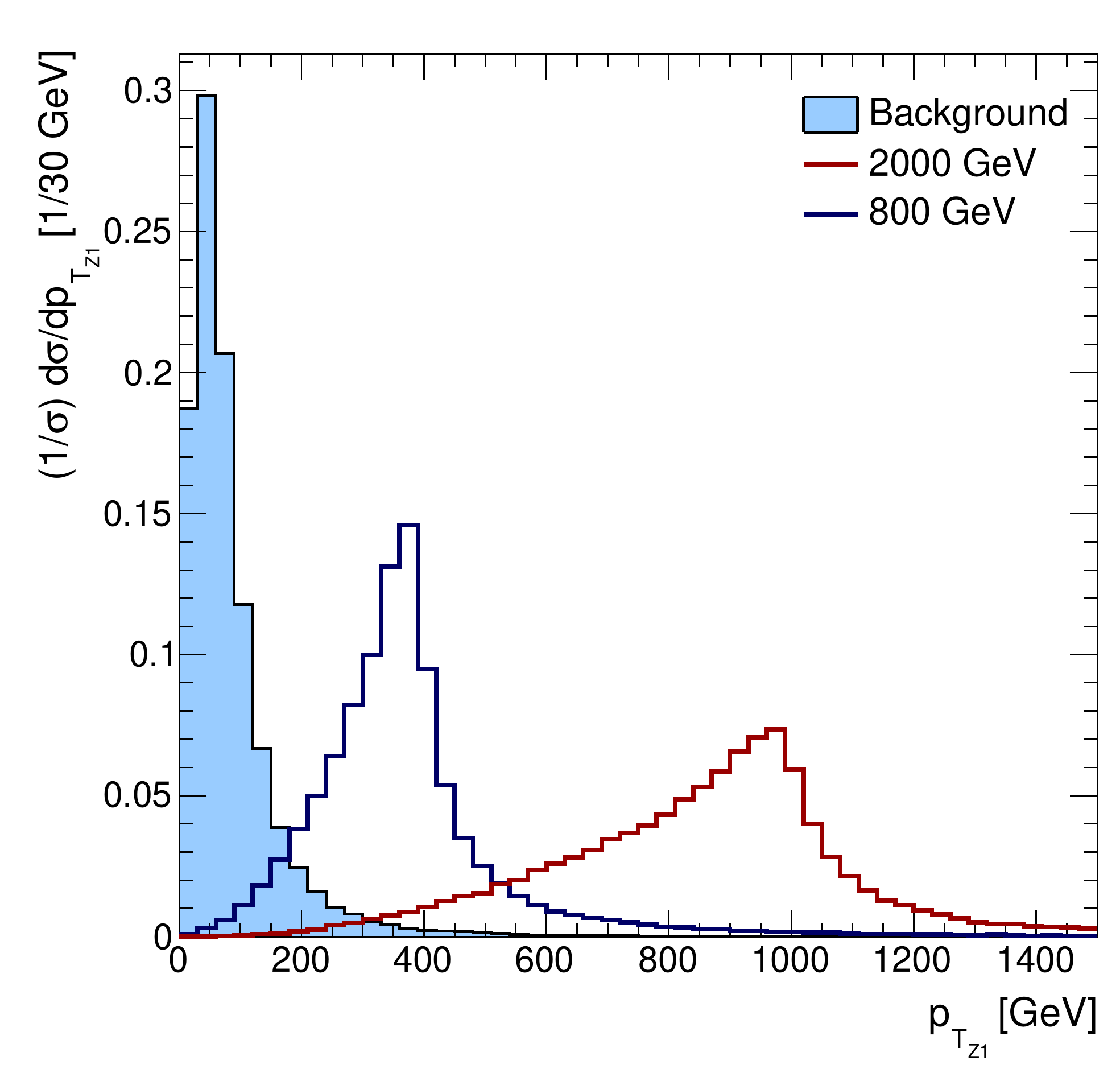}\label{fig:100TeV_Z1}}~
\subfloat{\includegraphics[scale=.20]{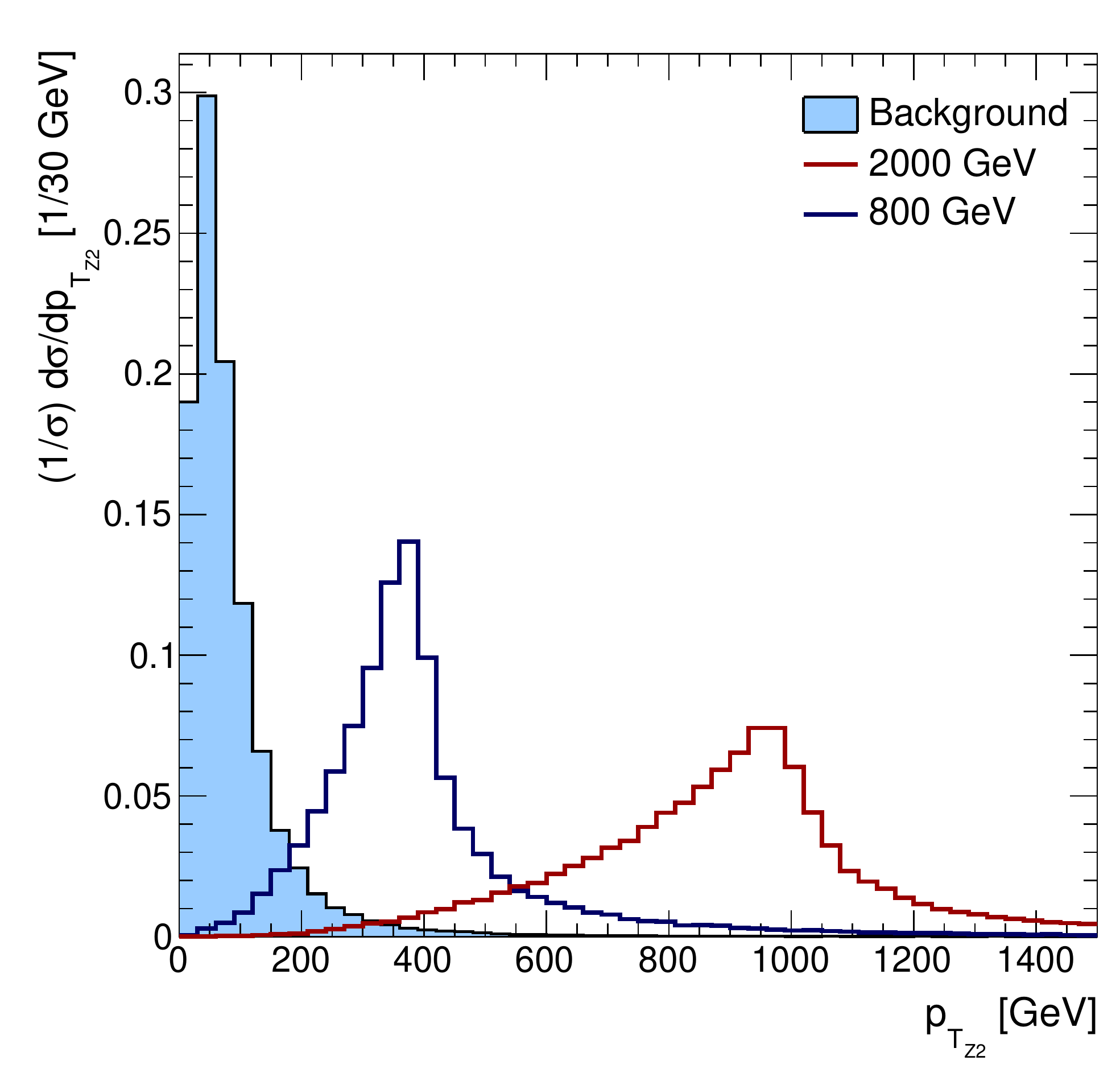}\label{fig:100TeV_Z2}}~\\
\subfloat{\includegraphics[scale=.20]{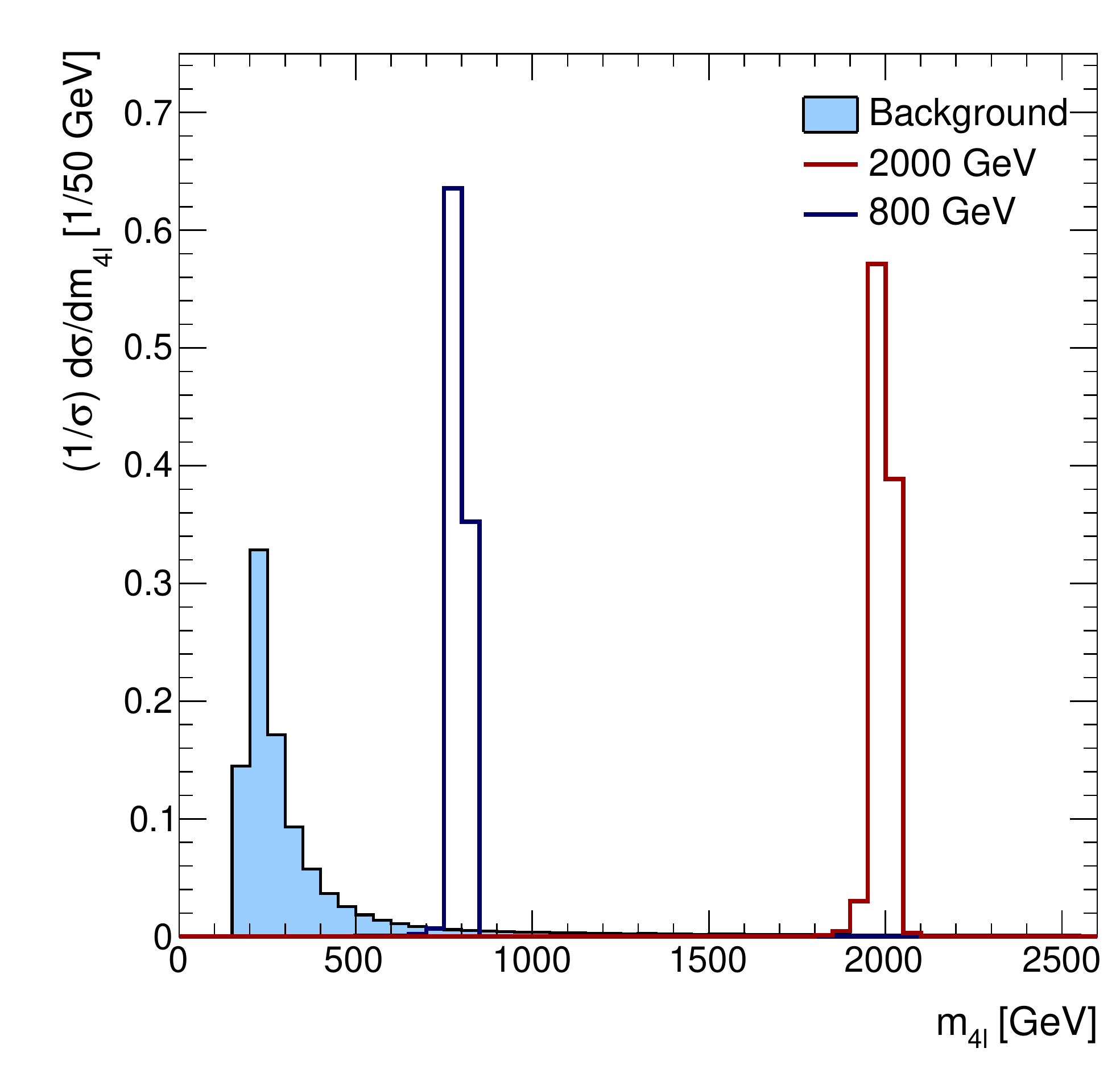}\label{fig:100TeV_m4l}}~
\subfloat{\includegraphics[scale=.20]{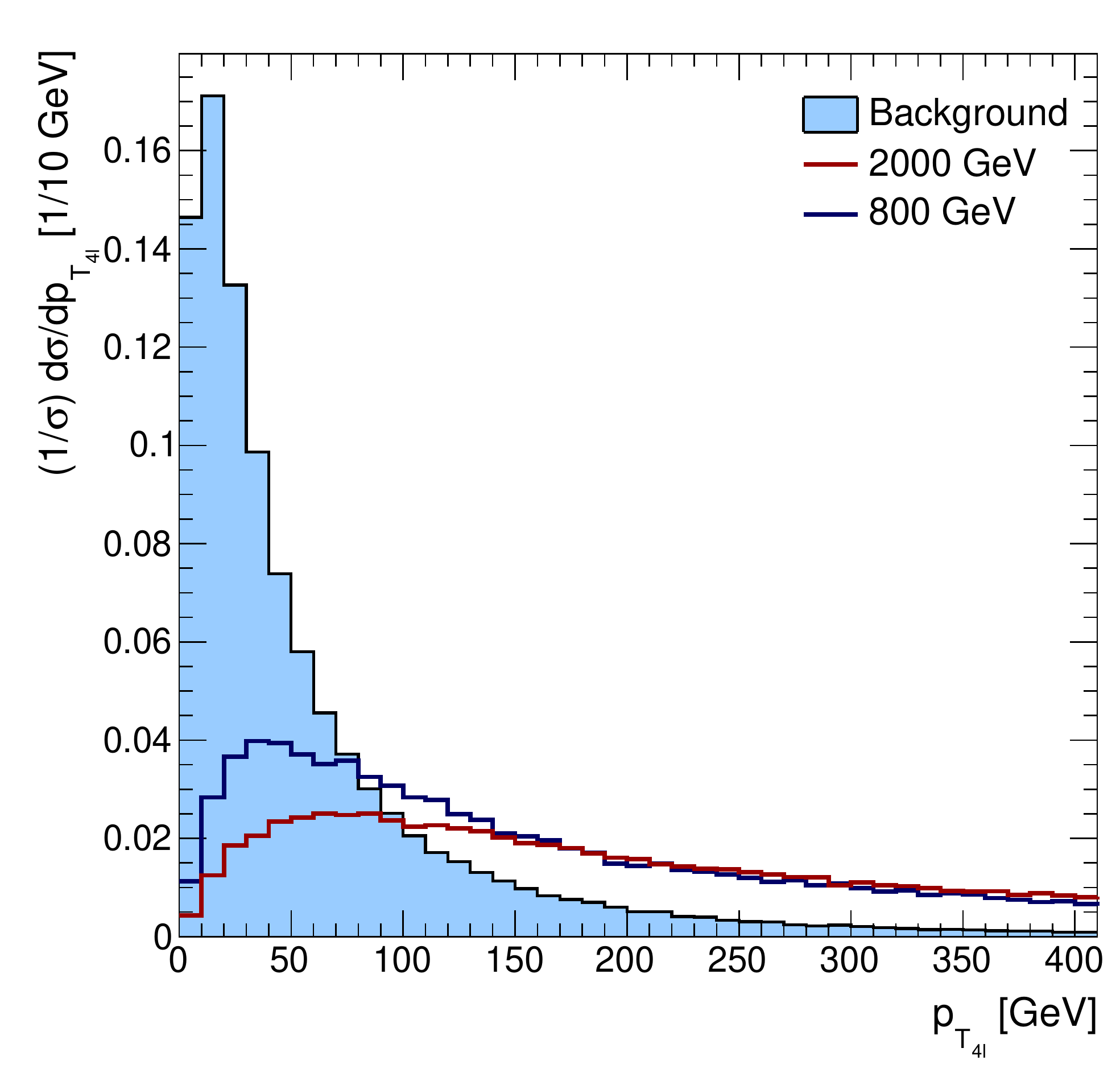}\label{fig:100TeV_pt4l}}
\caption{Normalized differential cross sections as a function of the transverse momentum of $\text{Z}_1$
boson (a) $\text{Z}_2$ boson (b), and the the invariant mass (c) and transverse momentum (d) 
of the four-lepton system for $m_{h_{2}}=2000$~GeV (red line), $800$ GeV (blue line) 
and background (blue solid) at $\sqrt{s}=100$~TeV.} 
\label{fig:Z100}
\end{center}
\end{figure}

\begin{figure}[htbp]
\centering
\subfloat{\includegraphics[scale=.20]{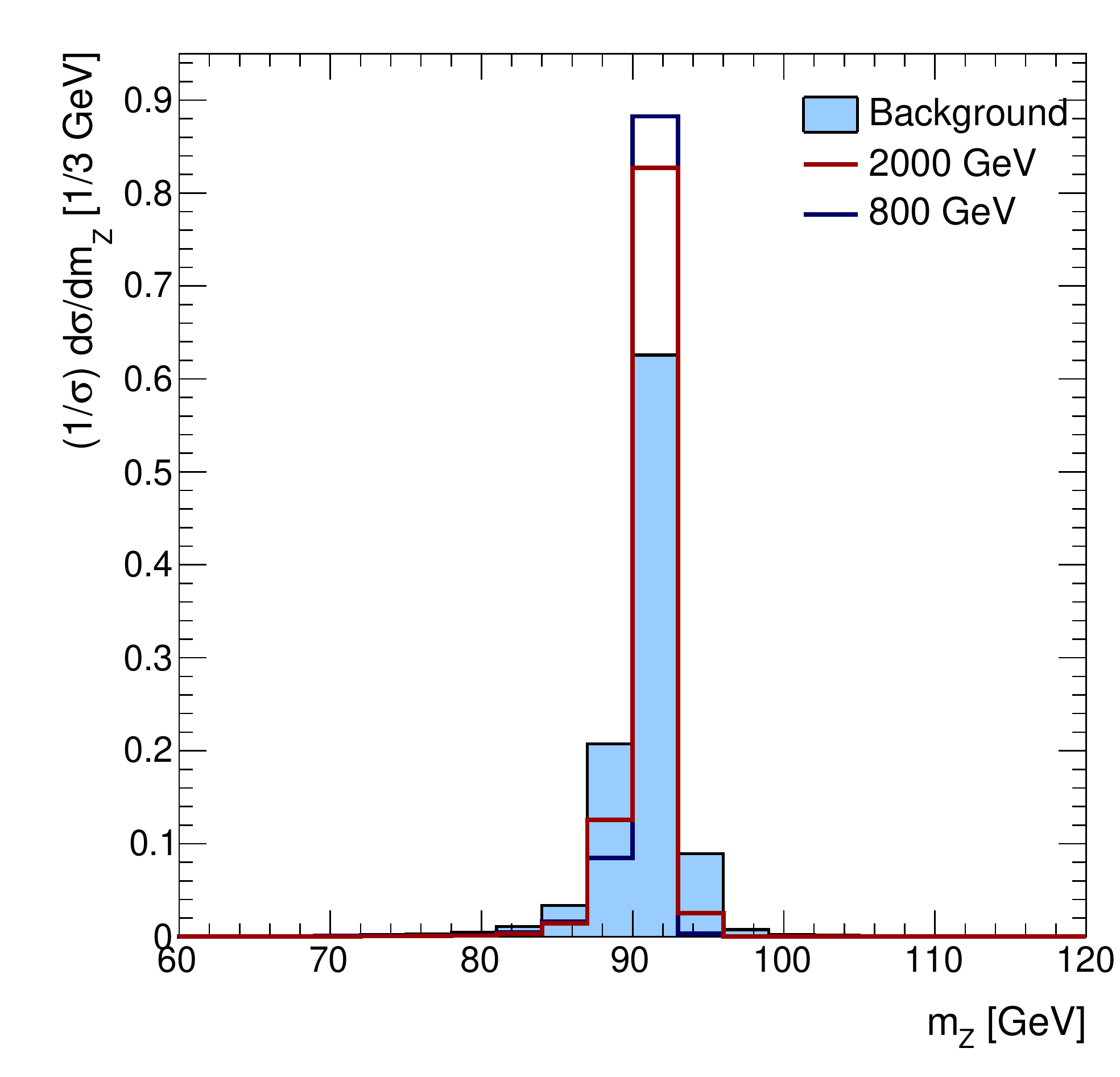}\label{fig:14TeV_Z}}~
\subfloat{\includegraphics[scale=.20]{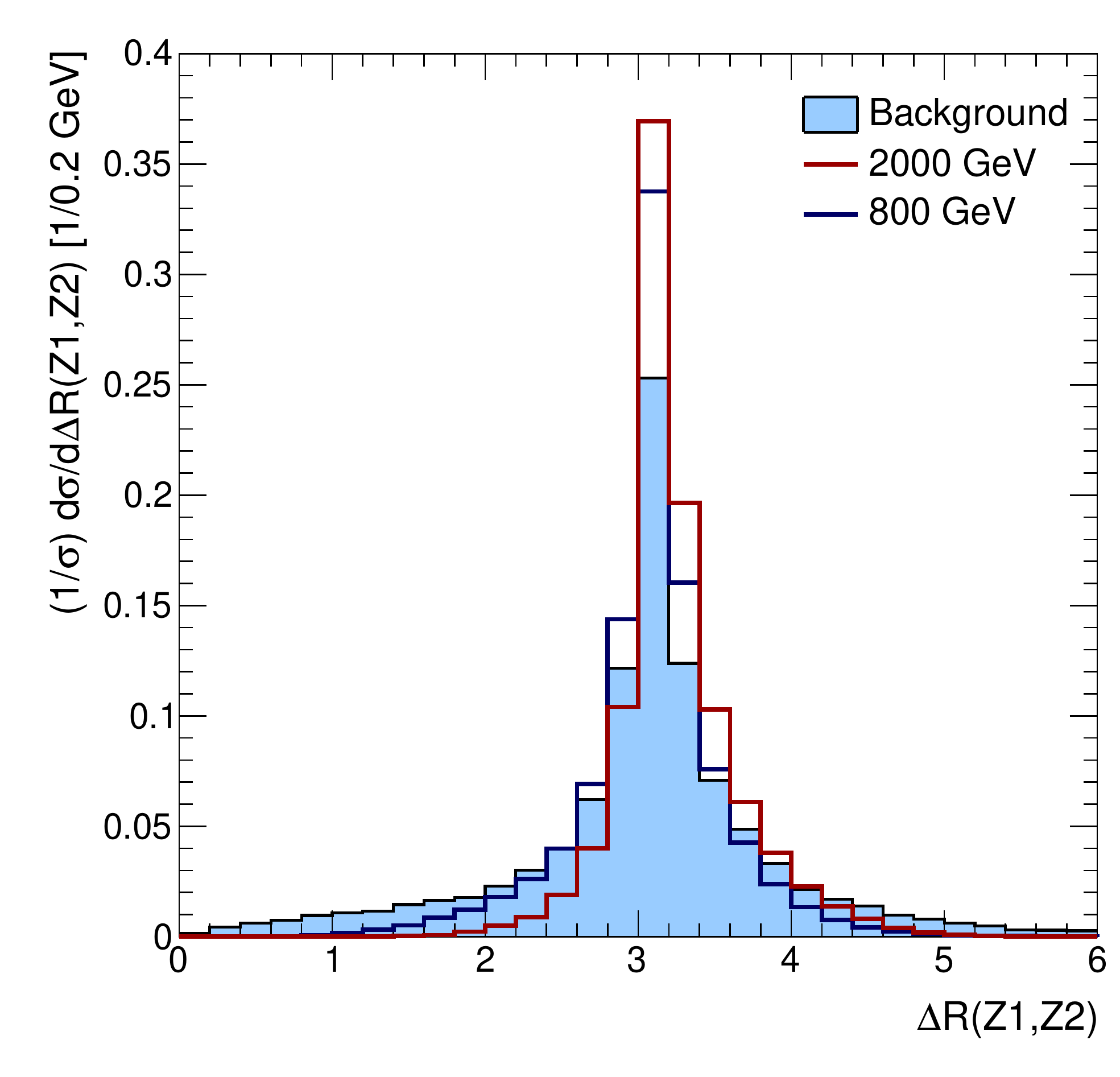}\label{fig:14TeV_drZZ}}~\\
\subfloat{\includegraphics[scale=.20]{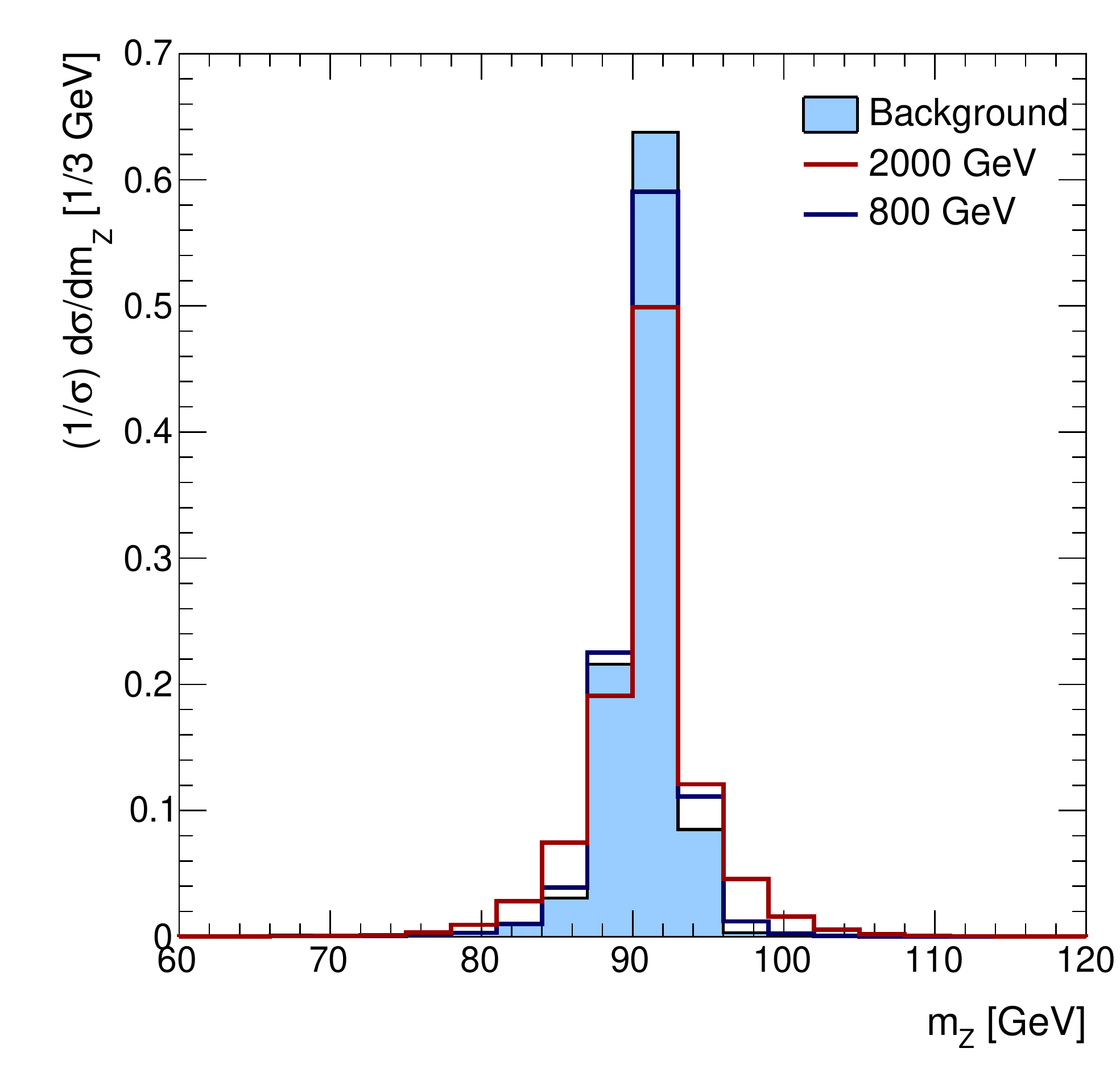}\label{fig:100TeV_Z}}~
\subfloat{\includegraphics[scale=.2042]{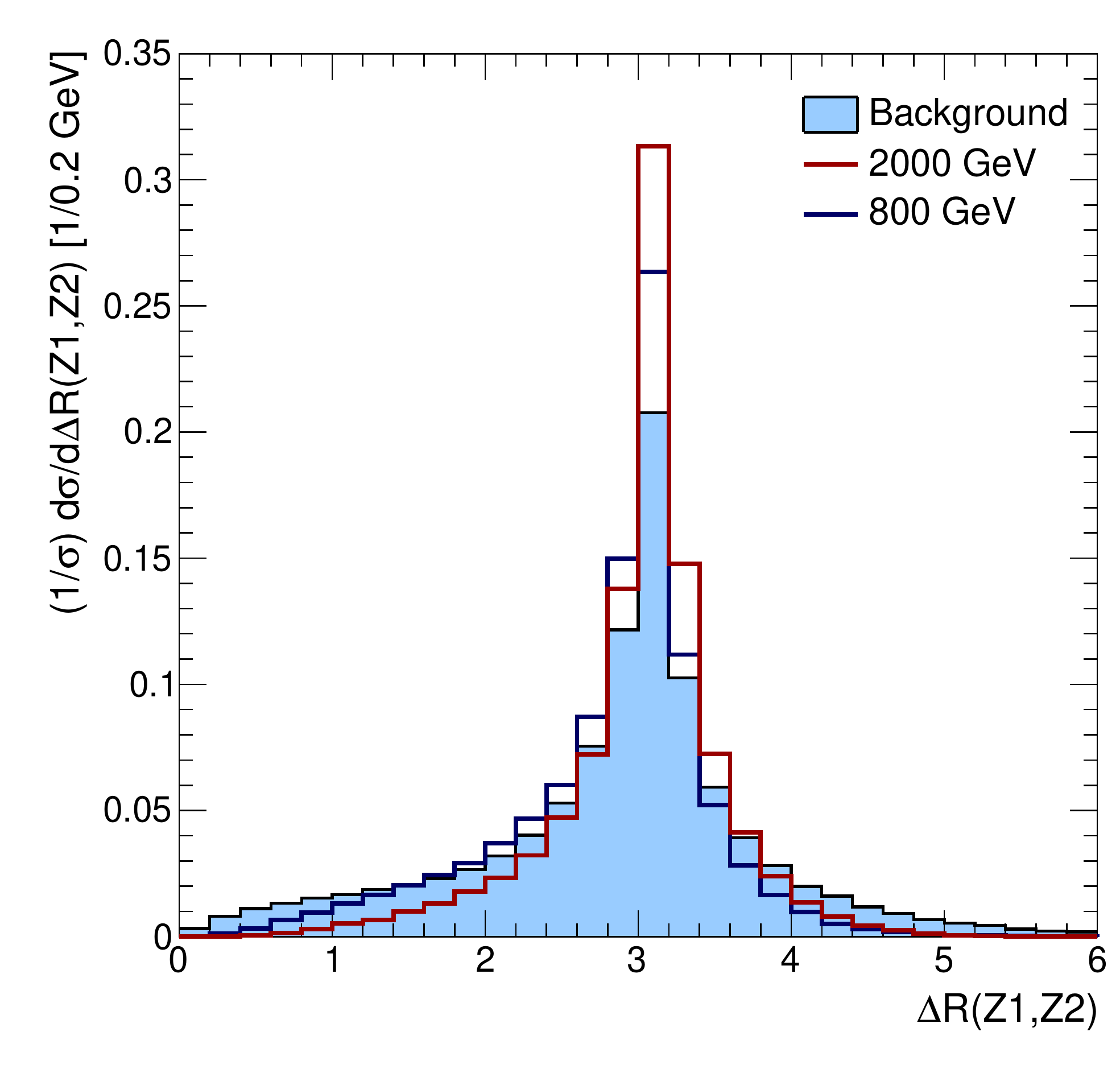}\label{fig:100TeV_drZZ}}
\caption{Invariant mass of the reconstructed Z boson candidates in the events (left) and 
angular separation ($\Delta \text{R} (\text{Z}_1, \text{Z}_2)$) between the two reconstructed Z bosons (right) at 14 TeV (up) 
and 100 TeV (down). The peak at $\sim 3$ in $\Delta \text{R} (\text{Z}_1, \text{Z}_2)$ distributions both for signal and 
background show that the Z bosons are produced mostly back-to-back.}
\label{fig:ZDeltaR}
\end{figure}

Depicted in Figs.~\ref{fig:lpt14}, \ref{fig:Z14}, 
\ref{fig:lpt100}, \ref{fig:Z100} and \ref{fig:ZDeltaR}
are the various kinematic distributions subjected
to the pre-selection cuts in Table \ref{Tab:precuts}.
Depicted in Fig.~\ref{fig:lpt14}, on the other hand, are the transverse momentum ($p_T$) spectra
of the leading, second-leading, third-leading, and fourth-leading lepton 
in signal events for $m_{h_{2}}=800$~GeV and $m_{h_{2}}=2000$~GeV at $\sqrt{s}=14$ TeV. The  
background contribution is also presented.
As shown in figure, the leptons originating from Z boson decays
in signal events (note that Z is the decay product of $m_{h_{2}}$) have harder $p_{T}$ spectrum 
while the leptons in background events significantly dominate in lower transverse momenta. 
The same distributions are given in Fig.~\ref{fig:lpt100} for the events at 100 TeV and the conclusion
is the same: leptons in signal events have harder $p_T$ spectra than the leptons in the background events. This feature makes lepton $p_T$ a decisive discriminator for extracting the signal from the background. 

Given in Fig.~\ref{fig:Z14} are the transverse momenta of the $\text{Z}_1$ and $\text{Z}_2$ bosons ($p_{T_{Z1}}$ and $p_{T_{Z2}}$) as well as the invariant mass ($m_{4l}$) and transverse momentum ($p_{T_{4l}}$) of the four-lepton system. The background is represented by the blue histograms.  The same distributions are shown in Fig.~\ref{fig:Z100} for $\sqrt{s}=$ 100 TeV. In similarity to the lepton $p_T$ spectra, the $p_T$ distributions of $\text{Z}_1$ and $\text{Z}_2$ bosons in the signal events, too, are seen to have  higher transverse momentum compared to those in the background.

It is clear from Fig.~\ref{fig:Z14} and Fig.~\ref{fig:Z100} that a cut of $p_{T}>120$ GeV on the transverse momenta of the both $\text{Z}_1$ and $\text{Z}_2$ bosons suppresses significant portion of background events.
Thus, it is possible to use also these distributions  to distinguish the signal from the background.
Moreover, the invariant mass of the four lepton system ($m_{4l}$) turns out to be one of the most sensitive 
observables in that it gives a narrow peak around the actual $m_{h_{2}}$ value, and it can thus be used to 
extract the signal especially when $m_{h_{2}}$ is large (for example, much larger than the $2m_\text{Z}$). (We do not use $p_T$ of the four-lepton system in event selection but we give them in Fig.~\ref{fig:14TeV_pt4l} and Fig.~\ref{fig:100TeV_pt4l}, for completeness.)

The Z-boson mass is a good measure of the success of the signal extraction. Shown in Fig.~\ref{fig:ZDeltaR} is the reconstruction of the two  Z boson candidates in the signal and background events at 14 and 100 TeV. This figure shows the invariant mass of the reconstructed Z candidates and 
$\Delta \text{R} (\text{Z}_{1}, \text{Z}_{2})$ distribution between the Z pairs. 
The reconstructed Z boson mass is seen to sharply peaked at the actual Z-mass $M_Z\sim 91.19\, {\rm GeV}$, ensuring that 
Z pairs in the events are reconstructed accurately enough.
The peak at $\Delta \text{R} (\text{Z}_{1}, \text{Z}_{2})\simeq 3$ reveals that the two Z bosons are mostly produced back-to-back.

For discriminating the signal from the background we apply a second set of selection criteria. Given in 
Table~\ref{Tab:cuts14} are the event selection and cumulative percentage efficiencies
after the sequential application of each cut, for the signal and background events at 14 TeV. The same data for 100 TeV samples is listed in Table~\ref{Tab:cuts100}.
In addition to Fig.~\ref{fig:ZDeltaR}, it can be seen also from these tables that the 
Z pairs in the events are reconstructed accurately enough. Despite the rather tight cut applied on the 
mass of the Z pairs ($\text{Z}_{1}$ and $\text{Z}_{2}$ $\in$ $\text{Z}$ $\pm$ 10 GeV), only approximately $4-5\%$ of the events are rejected, meaning that the Z bosons produced in the events are already within the $M_\text{Z} \pm 10\ {\rm GeV}$ mass range.
A large portion of the background signal is eliminated by the cut applied over the lepton momenta. 
Approximately 90\% of the background at 14 TeV and approximately 80\% at 100 TeV 
are eliminated by the cuts applied on the transverse momentum of the leptons in the final states.
The same cuts reject at most 15\% of the signal events at 800 GeV and 5\% at 2000 GeV.
The remaining background can be further suppressed with the additional cuts applied over the 
mass of the four lepton system ($m_{4l}$) and over the transverse momenta of the $\text{Z}_1$ and $\text{Z}_2$
bosons. However, compared to the signal processes, high production 
cross sections of the processes contributing to the background cause 
significant number of background events to arise at high luminosities. 
All these properties can be read off from Table~\ref{Tab:cuts14} (for 14 TeV) and Table~\ref{Tab:cuts100} (for 100 TeV).

\begin{table}
\caption{Effects of various selection cuts for the analysis of signal and background events at 14 TeV.
The values for the background are given for $m_{h_2}=800$~GeV by default while the values in parantheysis
indicate the remaining percentage of events for $m_{h_2}=2000$~GeV.}
\resizebox{\columnwidth}{!}{%
 \begin{tabular}{l|l|l|l}
 \hline
	Event selection  &   \begin{tabular}{@{}c@{}}Background \\ (\%) \end{tabular} & \begin{tabular}{@{}c@{}}$m_{h_{2}}=$ 800 GeV \\ (\%) \end{tabular} & \begin{tabular}{@{}c@{}}$m_{h_{2}}=$ 2000 GeV \\ (\%) \end{tabular} \\
 \hline
	\textbf{All:}                                                 & 100  & 100  & 100 \\
 \hline
	\textbf{Triggered:}\\
	number of pre-selected leptons $>=$ 4                                 & 16.3 & 32.6   & 36.8 \\
 \hline
 \hline
	\textbf{$\text{Z}$ reconstruction:}\\
	invariant mass of $m_{\text{Z}_1}$ and $m_{\text{Z}_2}$ $\in$ $m_\text{Z}$ $\pm$ 10 GeV/c$^2$  & 15.7 & 31.9   & 34.6 \\
 \hline
 \hline
	\textbf{Signal selection:} \\
	leading lepton $p_{T}>$ 90 GeV/c                          & 4.6 & 31.9   & 34.6 \\
	2nd leading lepton $p_{T}>$ 70 GeV/c                      & 3.4 & 31.8   & 34.6 \\
	3rd leading lepton $p_{T}>$ 50 GeV/c                      & 1.8 & 29.8   & 34.3 \\
	4th leading lepton $p_{T}>$ 20 GeV/c                      & 1.5 & 27.2   & 33.1 \\
	invariant mass of four lepton system \\ 
	($m_{4l}$) $\in$ $m_{h_{2}}$ $\pm$ 30 GeV/c$^2$             & 0.03 (0.0008) & 24.9   & 14.6 \\
	$p_{T_{Z_{1}}}>$ 120 GeV/c and $p_{T_{Z_{2}}}>$ 120 GeV/c & 0.03 (0.0008) & 24.6   & 14.6 \\
 \hline
 \hline
\end{tabular}
}
\label{Tab:cuts14}
\end{table}

\begin{table}
\caption{Effects of various selection cuts for the analysis of signal and background events at 100 TeV.
The values for the background are given for $m_{h_2}=800$~GeV by default while the values in parantheysis
indicate the remaining percentage of events for $m_{h_2}=2000$~GeV.}
\resizebox{\columnwidth}{!}{%
 \begin{tabular}{l|l|l|l}
 \hline
	Event selection  &   \begin{tabular}{@{}c@{}}Background \\ (\%) \end{tabular} & \begin{tabular}{@{}c@{}}$m_{h_{2}}=$ 800 GeV \\ (\%) \end{tabular} & \begin{tabular}{@{}c@{}}$m_{h_{2}}=$ 2000 GeV \\ (\%) \end{tabular} \\
 \hline
	\textbf{All:}                                                 & 100  & 100  & 100 \\
 \hline
	\textbf{Triggered:}\\
	number of pre-selected leptons $>=$ 4                                 & 7.8 & 34.9   & 44.3 \\
 \hline
 \hline
	\textbf{$\text{Z}$ reconstruction:}\\
	invariant mass of $m_{\text{Z}_1}$ and $m_{\text{Z}_2}$ $\in$ $m_\text{Z}$ $\pm$ 10 GeV$^2$    & 7.4 & 34.3   & 43.9 \\
 \hline
 \hline
	\textbf{Signal selection:} \\
	leading lepton $p_{T}>$ 90 GeV/c                      & 3.3 & 34.2   & 43.9 \\
	2nd leading lepton $p_{T}>$ 70 GeV/c                      & 2.7 & 34.1   & 43.9 \\
	3rd leading lepton $p_{T}>$ 50 GeV/c                      & 1.9 & 32.2   & 43.6 \\
	4th leading lepton $p_{T}>$ 20 GeV/c                      & 1.7 & 29.5   & 43.1 \\
	invariant mass of four lepton system \\ 
	($m_{4l}$) $\in$ $m_{h_{2}}$ $\pm$ 30 GeV/c$^2$             & 0.04 (0.01) & 29.0   & 37.4 \\
	$p_{T_{Z_{1}}}>$ 120 GeV/c and $p_{T_{Z_{2}}}>$ 120 GeV/c & 0.04 (0.01) & 28.4   & 37.4 \\
 \hline
 \hline
\end{tabular}
}
\label{Tab:cuts100}
\end{table}

After performing all the event selections and taking into account the expected total integrated 
luminosity values for the HL-LHC and FCC-hh, we compute the signal and background event yields as well as  the
signal significancies. The significancy is defined as
\begin{equation}
\label{eq:sign}
	n = \frac{S}{\sqrt{S+B}}
\end{equation}
where $S$ ($B$) is the number of signal (background) events which passed all the selection cuts.

\begin{figure}[htbp]
\begin{center}
\includegraphics[scale=.40]{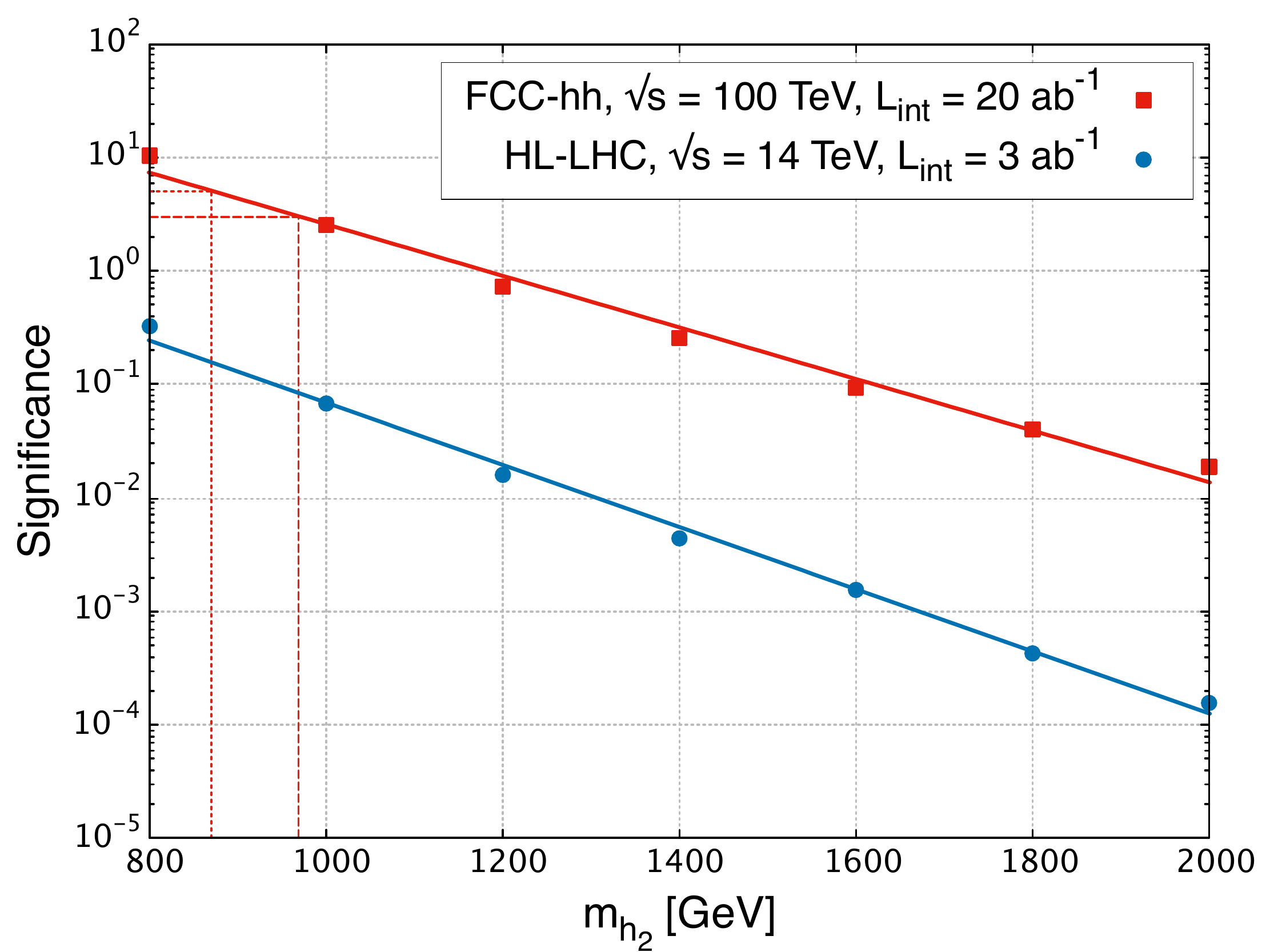}
\caption{Significance of the signals calculated as defined in (\ref{eq:sign}) as a function 
of $m_{h_{2}}$ for FCC-hh with an integrated luminosity
$\text{L}_{\text{int}}=$ 20~$\text{ab}^{-1}$ (a) and for HL-LHC with an integrated luminosity 
$\text{L}_{\text{int}}=$ 3~$\text{ab}^{-1}$.
The red dashed and red dotted lines indicate upper limits on $m_{h_{2}}$ for 3$\sigma$ evidence and 
5$\sigma$ discovery, respectively, for FCC-hh at $\text{L}_{\text{int}}=$~20~$\text{ab}^{-1}$.}
\label{fig:sig}
\end{center}
\end{figure}

The results obtained are plotted in Fig.~\ref{fig:sig} for FCC-hh and HL-LHC as a function of the $h_2$ mass.
The dots on the plot show the signal significance for the corresponding mass value
and the red and blue solid lines show the linear fits applied to the FCC-hh and HL-LHC values, respectively.
The red-dashed and red-dotted lines show, respectively, the upper limits on $m_{h_2}$ at 
3$\sigma$ (observation) and 5$\sigma$ (discovery).

It is clear that the resulting signal significances for the HL-LHC are too low: 0.3$\sigma$ at 800 GeV and gradually 
drops to 0.0001$\sigma$ at 2000 GeV. 
Therefore, it is clear that for the expected luminosity values of HL-LHC, which is 3~ab$^{-1}$, 
no significant excess of signal over the background is observed
in the 800~GeV$<m_{h_2}<$2000~GeV mass range. 
The results are more promising for FCC-hh. 
The higher center-of-mass energy of the collisions as well as the larger 
expected total integrated luminosity (100 TeV, L$_\text{int}=$ 20 ab$^{-1}$), 
enhance the signal significances by a factor of about 100.
At the FCC-hh, $h_2$ can be discovered
with a significance of 5$\sigma$ and a mass up to $\sim$870 GeV.
The upper limit on the mass for 3$\sigma$ evidence is $\sim$970 GeV. 
These limits can be pushed to higher values by using more 
advanced analysis techniques
with full simulation of the detectors when the FCC-hh begins to operate.

The signal significances are presented also in tabular form in Table~\ref{Tab:yields}
for each $m_{h_2}$ value considered in the analysis. (For FCC-hh, also event yields are shown.)
As $m_{h_2}$ increases, the signal significance gradually drops. This is true for both colliders. 
Higher luminosities would be needed to explore the multi-TeV mass region as 
the signal significance drops below 1$\sigma$ already at $\sim$1200 GeV for FCC-hh.  

\begin{table}[htbp]
\centering
\caption{Signal significances for HL-LHC and FCC-hh. For FCC-hh, also event yields are shown.}
 \begin{tabular}{l|l|l|l|l}
 \hline
                      & HL-LHC & \multicolumn{3}{c}{FCC-hh} \\
 \hline
    $m_{h_{2}}$ (GeV) &  $n$  &   $n$   &   S  &  B \\
 \hline
    800               &  0.326  & 10.36 &  778  & 4857 \\
    1000              &  0.067  & 2.533 &  154  & 3541 \\
    1200              &  0.016  & 0.733 &  40   & 2935 \\
    1400              &  0.004  & 0.254 &  12   & 2212 \\
    1600              &  0.001  & 0.093 &  4    & 1993 \\
    1800              &  4e-4   & 0.040 &  2    & 1612 \\
    2000              &  1e-4   & 0.019 &  0.6  & 1296 \\
 \hline
\end{tabular}
\label{Tab:yields}
\end{table}

\begin{figure}[htbp]
\begin{center}
\includegraphics[scale=.40]{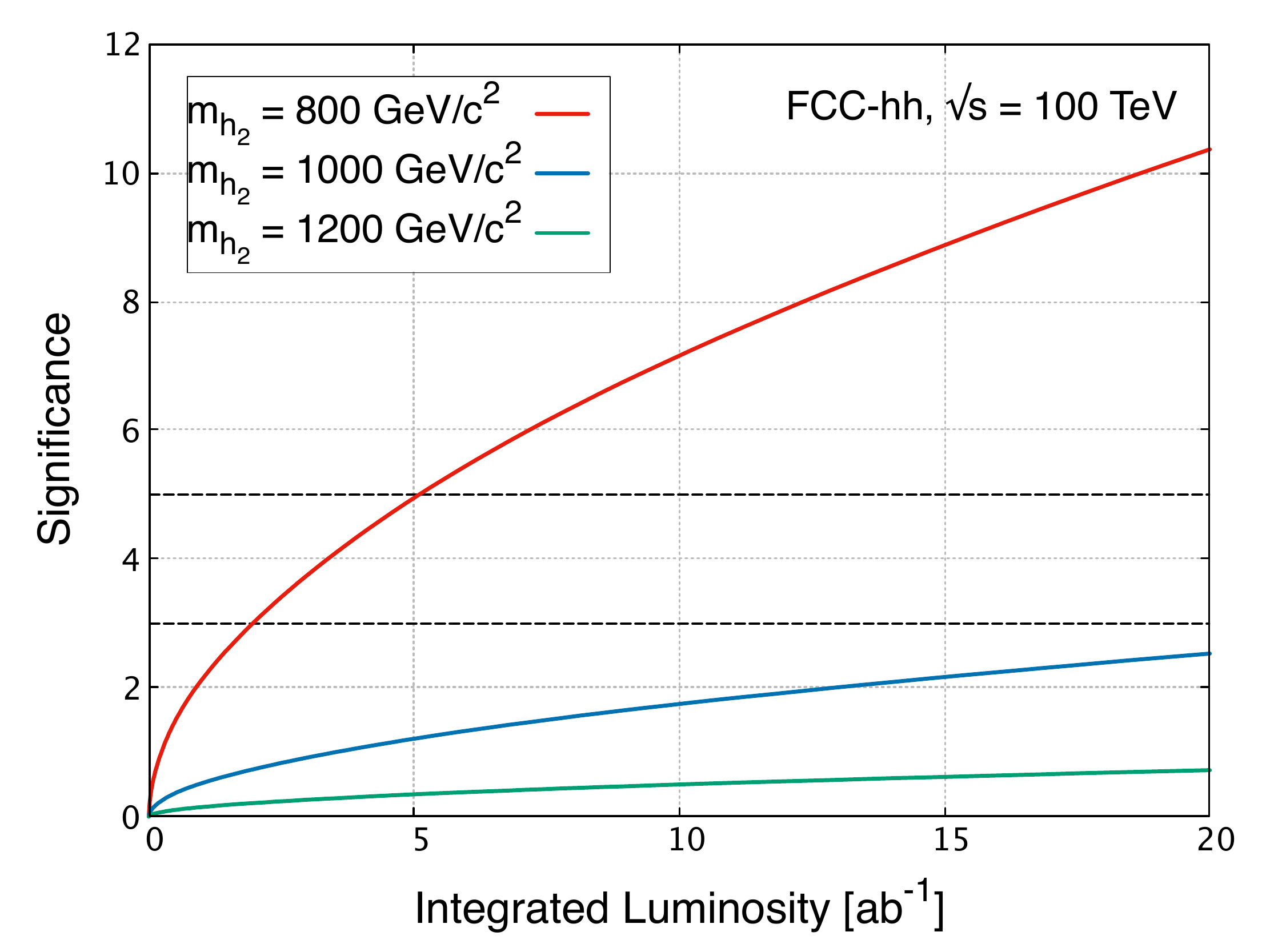}
\caption{Signal significance as a function of integrated luminosity at 100 TeV FCC-hh
for $m_{h_{2}}=800$, 1000 and 1200 GeV. The dashed lines correspond 3$\sigma$ and 5$\sigma$ values.}
\label{fig:siglumi}
\end{center}
\end{figure}

The signal significance as a function of integrated luminosity at 100 TeV FCC-hh for $m_{h_2}=$800, 1000, and 1200 GeV
is given in Fig.~\ref{fig:siglumi}. 
The results for the higher mass values are not shown since the values are 
too small. This figure ensures that if the mass of $h_2$ is about 800 GeV, the integrated luminosity required for a 5$\sigma$ discovery is 5.2 ab$^{-1}$. In addition, the results show that an evidence of $h_2$ with significance of 3$\sigma$
at the FCC-hh requires 2 ab$^{-1}$ integrated luminosity. This means that FCC-hh offers immediate discovery potential and
even at the early stages of its operating period,
we can start testing experimentally the BSM sector of symmergence. 

These simulation studies show that BSM models that do not destabilize the electroweak scale (such as symmergence) can be probed at relatiely  high luminosities as their couplings to the SM fields follow the seesawic scheme in (\ref{seesawic0}) and (\ref{linkup}).

\section{Conclusion}
In the present work we have studied impact of the electroweak stability on the collider discovery of the
BSM physics, with the example of a single SM-singlet scalar. The discussions in Sec.~\ref{sect:favored} and the simulation studies in Sec.~\ref{analysis} have revealed that fixing the SM-BSM coupling $\lambda_{HS}$ as in (\ref{lamhs}) (as a result of the MDDU structure in (\ref{linkup})) has important implications for new particle searches. This coupling fix, admissible in symmergence, tells us that there can exist heavy particles like $h_2$ and they can directly couple to the SM Higgs boson but they do not destabilize the electroweak scale thanks to see-sawic structure in (\ref{lamhs}). Though there is no apparent symmetry structure scalar field theories to support it, the empirical seesawic coupling (\ref{lamhs}) finds a rationale in the MDDU scheme in (\ref{linkup}).

The discovery potentials of $h_2$ in proton-proton collisions at the center-of-mass energies of 14 TeV and 100 TeV
are studied by analyzing $pp \to h_2 \to ZZ \to 4l$ channel,
with 3 ab$^{-1}$ total integrated luminosity for the HL-LHC and 20 ab$^{-1}$ for the future circular collider FCC-hh. Seven different mass scenarios for $h_2$, between 800-2000 GeV with increments of 200 GeV, are considered
for the searches. The detector effects are simulated via fast simulation package Delphes. Events with
two opposite-sign, same-flavor lepton pair in the final state are selected and the two Z boson candidates
are reconstructed event by event from the lepton pairs. Various observables, transverse momenta of the leptons,
mass and transverse momenta of the reconstructed Z bosons, and invariant mass of the four-lepton system,
are used to separate the signal from the background. The signal significances are calculated for each
mass value considered in the analysis.

For the HL-LHC with an expected integrated luminosity 3 ab$^{-1}$, 
no significant excess of signal over the background was observed
in the 800~GeV~$<m_{h_2}<$~2000~GeV mass range. 
The highest significance is 0.3$\sigma$ at 800~GeV
and gradually drops to 0.0001$\sigma$ at 2000 GeV. 
With higher collision energies and increased luminosities
at the FCC-hh (at 100 TeV with L$_\text{int}=$ 20 ab$^{-1}$), $h_2$ can be discovered with significance 
of 5$\sigma$ and a mass up to $\sim$870 GeV. If the mass of $h_2$ is about 800 GeV, 
the integrated luminosity required for a 5$\sigma$ discovery is 5.2 ab$^{-1}$.
In addition, the results show that an evidence for $h_2$ with significance of 3$\sigma$ 
at the FCC-hh requires 2 ab$^{-1}$ 
of integrated luminosity. This means that even at the early stages of the FCC-hh operating period,
we can start testing experimentally the BSM sector of symmergence. Considering 20 ab$^{-1}$,
the mass of $h_2$ can go up to a maximum of $\sim$970~GeV for a 3$\sigma$ evidence. 
Higher integrated luminosities or
searches for $h_2$ in different decay channels can push the limits forward, 
but this is beyond the scope of the analysis presented in this study.

It would be complementary to mention why lower mass values have not been analyzed. The main goal of the work is to stabilize the SM against heavy BSM fields and determine under what energy and luminosity ranges that heavy BSM can be discovered. With the example of an SM-singlet scalar, we have found that a seesawic coupling (supported by the MDDU linkup, as revealed in Sec. II B) does the job. By the nature of the seesawic coupling (which behaves as $\lambda_{HS}\leadsto \lambda_{SM}$ when $m_S^2 \leadsto m_H^2$), however, light scalars, $m_S^2 \leadsto m_H^2$, are expected to lead to SM-sized signal strengths. This low-$m_S$ domain is the one where the known SM completions like supersymmetry work, and the LHC data has already sidelined them for certain parameter regions \cite{Aad:2014vgg,Khachatryan:2015tra,Sirunyan:2017qfc,Sirunyan:2019tkw,Sirunyan:2019wph}. It is in this sense that our analyses have concentrated on heavy BSM rather than light BSM. 

Our analyses and discussions are based on the luminosities and center-of-mass energies of the existing
and planned (namely, the FCC) colliders. We have shown that the BSM (represented in this work by a 
single SM-single scalar $S$) can be probed properly in certain channels. It is clear that if these colliders
had larger luminosities we would be able to access higher-mass $h_2$ bosons at similar levels of significance.
Characteristically, electroweak stability necessitates seesawic couplings between the SM and the BSM, and seesawic couplings 
necessitate high luminosities for discovering the BSM. In domains where a BSM would be excluded
with ${\mathcal{O}}(1)$ couplings (as in supersymmetry, for instance) one can finds room for seesawically-coupled BSM. It is with much higher luminosities that colliders can access multi-TeV BSM, as exemplified by only ``TeV mass reach" of the planned FCC-hh collider. 

The discussions in the text have concentrated exclusively on the SM-singlet scalar $S$. The reason for this choice, as was mentioned in item (b) in Introduction, is that scalars form the  ``worst case" when it comes to electroweak stability since not only the Higgs mass but also the scalar mass are quadratically sensitive to the UV cutoff $\Lambda$. Their concurrent stabilization is possible in symmergence, and this has been utilized in the analyses in Sec. 2 and onward. But, a singlet scalar, though not strongly constrained by precision measurements and flavor physics, is difficult to search at colliders. To this end,  the hypercharge and lepton portals in (\ref{portals}) could be more promising.  In fact, electroweak stability is expected to allow for much heavier $Z^\prime$ boson and right-handed neutrinos $N$ when $|\lambda_{HS}| \sim |\lambda_{Z^\prime B}|\sim |\lambda_{LN}|$ since $Z^\prime$ and $N$ contributions to Higgs mass are quadratic in the couplings, as given in (\ref{portals}). Nevertheless, these rough estimates need be made precise by including all the available bounds. For instance, electroweak stability and known masses of active neutrinos together require the right-handed neutrinos to weigh below a 1000 TeV \cite{demir2019}. It thus follows that the present work need be extended to the $Z^\prime$ and $N$ sectors by incorporating available bounds (neutral currents, neutrino masses, flavor physics).  Their collider analyses, left to future work, can reveal important search strategies and distinctive signatures at future colliders like the LHC and FCC.

\section{Appendix}
\label{App}
The vertex factors in (\ref{eq:logcontr}) have the following expressions:

\begin{multline}\label{eq:l1111}
\lambda_{h_1h_1 h_1h_1} =\frac{\lambda_H}{4}\cos^4\theta+\frac{\lambda_S}{4}\sin^4\theta+\frac{\lambda_{HS}}{16}\sin^22\theta\nonumber
\end{multline}
\begin{multline}
\lambda_{h_1h_1\phi\phi}=\lambda_{h_1h_1\phi_0\phi_0}=\lambda_{h_1h_1\phi_1\phi_1}=\lambda_{h_1h_1\phi_2\phi_2}\nonumber\\
=\frac{\lambda_H}{2}\cos^2\theta+\frac{\lambda_{HS}}{4}\sin^2\theta
\end{multline}
\begin{multline}
\lambda_{h_1h_1h_2h_2}=\frac{3}{8}\left(\lambda_H+\lambda_S\right)\sin^22\theta\\
+\frac{\lambda_{HS}}{4}\left(\cos^4\theta+\sin^4\theta-\sin^22\theta \right)\nonumber
\end{multline}
\begin{multline}
\lambda_{h_1h_1h_1}=\lambda_H\upsilon_H\cos^3\theta-\lambda_S\upsilon_S\sin^3\theta\\
-\frac{\lambda_{HS}}{4}\left(\cos\theta \upsilon_S-\sin\theta \upsilon_H\right)\sin2\theta\nonumber
\end{multline}
\begin{multline}
\lambda_{h_1\phi\phi}=\lambda_{h_1\phi_0\phi_0}=\lambda_{h_1\phi_1\phi_1}=\lambda_{h_1\phi_2\phi_2}\\
=\lambda_H\upsilon_H\cos\theta-\frac{\lambda_{HS}}{2}\upsilon_S\sin\theta \nonumber
\end{multline}
\begin{multline}
\lambda_{h_1h_2h_2}=\frac{3}{2}\left(\lambda_H\upsilon_H\sin\theta-\lambda_S\upsilon_S\cos\theta\right)\sin2\theta\\
+\frac{\lambda_{HS}}{2}\left(\left(\cos\theta\sin2\theta-\sin^3\theta\right)\upsilon_S \right. \nonumber\\
\left.+\left(\cos^3\theta-
\sin\theta\sin2\theta\right)\upsilon_H\right)\nonumber
\end{multline}
\begin{multline}
\lambda_{h_1h_1h_2}=\frac{3}{2}\left(\lambda_H\upsilon_H\cos\theta+\lambda_S\upsilon_S\sin\theta\right)\sin2\theta\\
+\frac{\lambda_{HS}}{2}\left(\left(\cos^3\theta-\sin\theta\sin2\theta\right)\upsilon_S\right.\nonumber\\
\left.+\left(\sin^3\theta-\cos\theta\sin2\theta\right)\upsilon_H\right)
\end{multline}

\bigskip
\noindent\textbf{Acknowledgments}\\
This work is supported in part by the T{\"U}B{\.I}TAK grant 118F387 and by the {\.I}TU BAP  grant TAB-2020-42312. We thank Beyhan Puli{\c c}e for her contributions at the initial stage of this work. We are grateful to conscientious referee for her/his constructive comments, criticisms and suggestions.  
%

\end{document}